\documentclass[aip,reprint,floatfix,superscriptaddress]{revtex4-2}

\usepackage{amsmath}
\usepackage{amssymb}
\usepackage{accents}
\usepackage{tabularx}
\usepackage{graphicx}
\usepackage{rotating}
\usepackage{dcolumn}
\usepackage{xr}
\usepackage{cleveref}
\usepackage{bm}
\usepackage{xcolor}
\UseRawInputEncoding

\begin{document}

\date{\today}\title{Electrothermal Actuation of NEMS Resonators:\\  Modeling and Experimental Validation}

\newcommand{\BU}{Department of Mechanical Engineering, Division of Materials Science and Engineering, and the Photonics Center, Boston University, Boston, Massachusetts 02215, United States}

\author{Monan Ma}
\email[Electronic mail: ]{monan@bu.edu}
\affiliation{\BU}

\author{K. L. Ekinci}
\email[Electronic mail: ]{ekinci@bu.edu}
\affiliation{\BU}

\date{\today}

\begin{abstract}
 
We study the electrothermal actuation of nanomechanical motion using a combination of numerical simulations and analytical solutions. The nanoelectrothermal actuator structure is a u-shaped gold nanoresistor that is patterned on the anchor of a doubly-clamped nanomechanical beam or a microcantilever resonator. This design has been used in recent experiments successfully. In our finite-element analysis (FEA) based model, our input is an ac current; we first calculate the temperature oscillations due to Joule heating using Ohm's Law and the heat equation;  we then determine the thermally induced bending moment  and the displacement profile of the beam by coupling  the temperature field to Euler-Bernoulli beam theory with tension. Our model efficiently combines transient and frequency-domain analyses: we compute the temperature field using a transient approach and then impose this temperature field as a harmonic perturbation for determining the mechanical response in the frequency domain. This unique modeling method offers lower computational complexity and improved accuracy, and is faster than a fully transient FEA approach. Our dynamical model computes the temperature and displacement fields in time domain over a broad range of actuation frequencies and amplitudes. We validate the numerical results by directly comparing them with experimentally measured displacement amplitudes of NEMS beams around their eigenmodes in vacuum. Our model predicts a thermal time constant of 1.9 $\rm ns$ in vacuum for our particular structures,  indicating that electrothermal actuation is efficient up to $\sim80$ MHz. We also investigate the thermal response of the actuator when immersed in a variety of fluids. 
\end{abstract}

\maketitle

\section{Introduction}
An electromechanical actuator,\cite{kouh2017nanomechanical, Ekinci2005small} in the most general sense, converts a given signal from the electrical domain into the mechanical domain. This conversion is accomplished by applying a time-dependent force to a mechanical element, which causes prescribed movements. In many applications, the actuator drives the mechanical element by a harmonic force, causing it to oscillate. More complicated mechanical signals in a variety of applications---from macroscopic robots \cite{Liang2012,Jin2020,pop2022lithium} to mechanical computers \cite{Mahboob2008,Qian2017}---are accomplished by a  host of different electromechanical actuators, working at different length and time scales and based on different physical actuation principles. The current trend  for creating electromechanical systems with micrometer- and nanometer-scale linear dimensions has necessitated the development of reliable and efficient miniaturized actuators. Thus, in the fields of  micro- and nano-electro-mechanical  systems (MEMS \cite{maluf2004introduction,villanueva2008crystalline,chui1998independent} and NEMS \cite{kouh2017nanomechanical,mile2010plane,li2007ultra,bargatin2007efficient,tang2009metallic}), electromechanical actuators have been one of the research foci. Some recent areas of interest in miniaturized actuator research are: exploring novel physical mechanisms \cite{Nan2013,Huang2005,Unterreithmeier2009} and exotic materials \cite{Feng2007,Masmanidis2005} for nanoscale actuators; developing ultra-responsive actuators \cite{Li2017}  that work reliably in different environments, including in liquids \cite{Andreassen2014,ti2022,ari2020nanomechanical} and harsh environments;\cite{Singh2014,Rajaram2022} and formulating accurate physical models of actuators.\cite{Koochi2015,Yazdanpanahi2014,Farrokhabadi2016,Ouakad2017,liem2020inverse,Barbish2022} 
 
Different actuation approaches based on electrical and optical coupling have been explored in the domain of NEMS. Here, we provide a brief---and admittedly incomplete---review of the mainstays in the electrical domain; more details can be found in recent review articles.\cite{kouh2017nanomechanical,Hajjaj2020,Khaniki2021} Using electrical coupling, appreciable forces can be exerted on MEMS and NEMS devices through piezoelectric,\cite{Mahboob2008, Masmanidis2007,Wasisto2014} electrostatic, \cite{truitt2006, mile2010plane} and electrothermal \cite{jiang2006sic,reichenbach2006,bargatin2007efficient} transducers. Most of these actuation methods have been modeled and studied parametrically. Modeling of electrically-coupled actuators typically requires numerical simulations, although analytical models are also useful for simpler structures. 
\begin{table*}[t]
\caption{Comparison of modeling techniques for electrothermal actuators published within the last 5 years.} \label{comparison}
\setlength{\tabcolsep}{11pt}
\renewcommand{\arraystretch}{1.2}
\begin{ruledtabular}
\begin{tabular}{lccc}
Actuator type & Model approach  & Validation method & References \\
\hline
Multilayer & FEA quasi-static &  Experimental & Ref. \cite{ren2021electro}\\
Hot-arm  & FEA quasi-static & Analytical & Ref. \cite{cauchi2018analytical, masood2019design} \\
Chevron & Analytical quasi-static & FEA & Ref. \cite{thangavel2018modelling, hussein2020modeling}\\
Bimorph & Analytical quasi-static &  Experimental & Ref. \cite{kim2022analytical}\\
Bimorph & FEA dynamic &  Experimental & this work \\
\end{tabular}
\end{ruledtabular}
\end{table*}

In the case of piezoelectric actuation, an applied electric field results in a mechanical stress in the material. Inversely, a mechanical strain in a piezoelectric material results in a proportional electrical field, which can be detected and related to the amplitude of the displacement. Typical piezoelectric materials used for manufacturing nanoscale actuators and resonators include AlN \cite{Cleland2001}, GaAs \cite{Knobel2002}, and AlGaAs \cite{Adachi1985}. Coupling between mechanical and electrical domains of a piezoelectric actuator is governed by linear piezoelastic relations, \cite{piezoelastic} and the stress and strain can be computed from the electric field and polarization density and \textit{vice versa}. In a study of small displacements of a doubly-clamped nanomechanical beam, Knobel and Cleland \cite{Knobel2002} simplified the constitutive relations into a lumped model, where the displacement-induced electrical charge was calculated from Ohm's law for a capacitor and the lumped force was related to the beam antinode displacement through Hooke's law.  In another study by Sinha \textit{et al.}, \cite{Sinha2009} two layers of ultrathin piezoelectric AlN were stacked in a NEMS cantilever for inducing  vertical bending through the bimorph effect. A finite-element software was employed to simulate the displacement amplitude of the cantilever for various geometries and applied voltage values, and closely matched the experiments. 

For electrostatic (capacitive) actuators, the interaction force is simply governed by Coulomb's law. In the domain of MEMS and NEMS, the magnitude of the elastic force becomes comparable to the electrostatic force, and the displacement can be computed from the balance of forces. At the sub-micron length scale, the electrostatic actuator cannot be assumed to be two ideal capacitor plates. \cite{schwab2007capacitive} The plate dimensions become comparable to the separation distance, and the contributions from the sides of the structure become significant, resulting in fringe effects. \cite{Bao2005MEMSbook} Furthermore, as the electrostatic force varies non-linearly with the separation distance and due to the presence of geometric nonlinearities in the structure, it is challenging to precisely compute the force balance. As a result, it is generally impossible to come up with a closed-form solution for a micro- or nanoscale capacitive actuator system, \cite{Bao2005MEMSbook} and numerical approaches are preferred. \cite{younis2011} Shavezipur \textit{et al.} \cite{Shavezipur2011} have performed a finite-element study using Poisson's equation and relevant boundary conditions for a MEMS capacitive switch, where the contact was established by an electrostatic force. Their numerical model accurately captures the electrostatic force and capacitance, whereas a simplified analytical model, which neglects fringe effects, will under-predict these values.

In this manuscript, we focus on electrothermal actuation of NEMS resonators. \cite{bargatin2007efficient, jiang2006sic} Electrothermal actuation is based on a nanoscale distributed resistor fabricated on the anchor of the NEMS structure. The actuation mechanism is based on Joule heating from a supplied current, whereby the temperature gradients induce  bending moments via thermoelastic processes. This type of actuation can produce large forces for relatively low drive currents \cite{mao2010, zhu2021} and provide a large bandwidth of actuation, thus allowing for many harmonic modes to be actuated efficiently. Moreover, electrothermal actuators can operate in liquids, \cite{ari2020nanomechanical,ti2022} and are compatible with complementary metal-oxide-semiconductor (CMOS \cite{goldsmith1999, parameswaran1990cmos, reinke2010}) technology. Other well-established designs of micro- and nanoscale electrothermal actuators include hot-arm actuators, \cite{Castro2017, Elbuken2009, Yan2003, Huang1999} where the actuator system consists of a more electrically resistive hot arm and a less resistive cold arm. This resistance mismatch results in asymmetric Joule heating, which in turn generates mechanical displacement. Another configuration is a V-shaped chevron actuator, \cite{Shan2017, Enikov2005} where a linear displacement can be achieved by Joule heating-induced thermal expansion of symmetric connecting arms. 

Previous models of electrothermal actuators mainly focus on the quasi-static mechanical response due to electrical drive. The analytical thermal model was developed based on Joule heating and heat transfer. For slender actuators, thermal equations can be simplified further by approximating the actuator as a 1-D system, \textit{i.e.,} the temperature is the same for the entire cross-sectional area at a particular position along the actuator. The calculated temperature distribution is then related to mechanical displacements and bending moments using relevant constraining boundary conditions, linear thermal expansion, and the virtual work method. \cite{Yan2003, Elbuken2009, Huang1999} The results from analytical equations agree well with finite-element simulations and experiments, but do not provide insight into the dynamic response of a system to a periodic excitation. To this end, Bargatin \textit{et al.} \cite{bargatin2007efficient} simulated and experimentally validated the electrothermal actuation efficiency as a function of the drive frequency for a doubly-clamped NEMS resonator. The resonator's frequency response resembled a low-pass filter partly due to the effect of thermal roll off, where at high drive frequencies the temperature increase lags behind the Joule heating, resulting in a smaller  temperature oscillation. In another numerical study, Hickey \textit{et al.} \cite{Hickey2003} investigated time and frequency response of two-arm electrothermal actuators. Here, the thermal model was based on the one-dimensional Joule heating, and the model captured the contribution from all three modes of heat transfer, which allowed for computing the thermal time constant and thermal response at various drive amplitudes and frequencies. However, the model did not explicitly show the subsequent mechanical response of the system. As a result, a comprehensive, dynamical model of oscillations that reveals essential variables of interest, such as the instantaneous bending moment, actuator temperature field and the mechanical displacement field across a broad range of excitation frequencies, has not yet been developed. This is primarily due to the intricacy of the coupled physical phenomena, which are sensitive to the specifics of the model.  Understanding the thermal and mechanical responses of a system is vital to designing an efficient electrothermal actuator and predicting its actuation limits both in terms of frequency bandwidth and power dissipation. We present a comparative table in order to put our modeling approach into context with respect to the alternative analytical and FEA models developed within the last five years. As shown in Table \ref{comparison}, we make comparisons across the types of electrothermal actuators, the modeling approaches and the validation methods.

Our aim in this manuscript is to develop an extensive and reliable model of electrothermal actuators by efficiently combining relevant physics domains. We first compute the steady-state oscillations of the temperature field in a NEMS resonator in response to input current at a particular excitation frequency due to Joule heating. We then calculate the displacement field of the  resonator by considering the effects of differential thermal expansion of the structure. We validate our model by comparing numerical results with recent experimental measurements from literature: our model matches both the frequency and the amplitude observed in the experiments accurately.  We also calculate the thermal stresses and the induced bending moments in the structure. Finally, we discuss how these numerical results can also be incorporated into the Euler-Bernoulli beam theory with tension. \cite{Barbish2022} The structure of this manuscript is as follows. In Section II, we introduce the NEMS geometry under study and describe the details of the numerical approach. In Section III, we present the results of FEA simulations and compare these with experiments. We also study the thermal response of the structure when immersed in a fluid. Finally, in Section IV, we discuss our numerical approach and its connection with Euler-Bernoulli beam theory with tension.

\section{Approach}

\subsection{Device Geometry and Parameters}
\begin{figure}
    \includegraphics{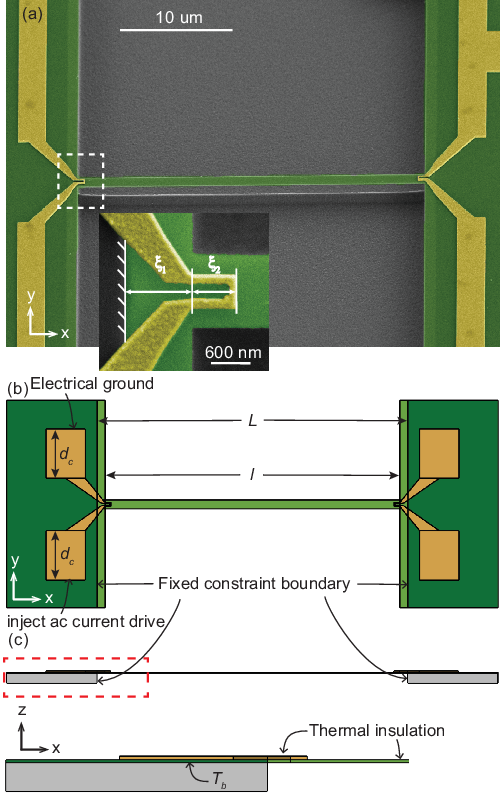}
    \caption{(a) False-colored SEM image of a silicon nitride (green) doubly-clamped beam with u-shaped gold nanoresistors (yellow) patterned on both anchors. The linear dimensions of the beam are $l\times w \times h \approx \rm 30~\mu m \times 900~nm \times 100~nm$. The bright green region is the suspended part of the device, which includes the beam and two undercuts. The inset shows a magnified SEM image of the nanoresistor, \textit{i.e.,} the electrothermal actuator, along with the suspended part of the SiN layer and the location of the clamping surface. Nanoresistor cross-section is $\rm 120~nm \times 135~nm$; the suspended (undercut) region on the beam is $\xi_1 \approx 800 \rm ~nm$, and the region on the anchor is $\xi_2 \approx 600 \rm ~nm$. (b) Top view of the full geometry used for the FEA simulations with imposed electrical and structural boundary conditions. (c) Side view of the model focusing on the electrothermal actuator with applied thermal boundary conditions.}
    \label{Figure1}
\end{figure}
Our study is based on commonly-used NEMS doubly-clamped beam resonators with tension patterned on a silicon nitride (SiN) wafer. Scanning electron microscope (SEM) image of a representative beam is shown in Fig. \ref{Figure1}(a). The linear dimensions of the beam are $l\times w \times h \approx \rm 30~\mu m \times 900~nm \times 100~nm$. There is an undercut region at both ends of the beam with a length of $\xi_1\approx800~\rm nm$. Hence, the beam itself and the two undercut regions make up the suspended part of the device. We denote the total length of the suspended part of the device as $L$, and thus $L = l + 2\xi_1$, as shown in Fig. \ref{Figure1}(b). The suspended region is shown in bright green in the SEM image in Fig. \ref{Figure1}(a) and illustrated in Fig. \ref{Figure1}(b).  The boundary line at the interface between the bright and dark green regions is assumed to be clamped to the substrate underneath, as indicated in the inset of Fig. \ref{Figure1}(a). The inset also shows the magnified image of the gold nanoresistor, which is the electrothermal actuator. The nanoresistor and the connecting pads are patterned using electron beam lithography. Each nanoresistor has a thickness of $135 \rm ~nm$ and a width of $120 \rm~nm$; $\xi_1\approx 800~\rm nm$ and $\xi_2 \approx 600~ \rm nm$ are respectively the lengths of the nanoresistor fabricated on top of the undercut and the beam. The following relevant parameters were determined in previous experiments: \cite{chaoyang2021} the intrinsic tension in all the beams is $F_T = 68.8 \pm 12.9 \rm ~\mu N$; the Young's modulus of SiN is  $E_{s} \approx 300 \rm ~GPa$; and the density of the SiN is $\rho_{s} = 2960 \pm 30 \rm ~kg/m^3$. The resistivity of the gold film is $\rho_{g} \approx 2.82 \times 10^{-8} ~{\Omega} \cdot \rm m$, and the total resistance of the nanoactuator, including the contacts, is $R \approx 25~\Omega$.\cite{Ticomm}

In order to enable a direct comparison between the finite-element model and experiments, we define a set of model parameters and boundary conditions. Specifically, we numerically compute the beam's intrinsic stress $s_{xx}^{(0)}$ and modal isotropic dissipation $\eta_n$ so that the modal frequencies $f_n$ and quality factors $Q_n$ correspond to experimental measurements. The computed intrinsic tensions used for simulation, \textit{i.e.,} $s_{xx}^{(0)}/(wt)$, are $75.3~\mu \rm N$ and $73.4~ \mu \rm N$ for the $30$-$\mu \rm m$ and $50$-$\mu \rm m$ beams, respectively, which agrees with the experimentally measured tension force $F_T$ for high-tension beams.\cite{ari2020nanomechanical} The main factors affecting the $Q_n$ of the devices are intrinsic losses, \textit{e.g.,} defects, and clamping losses;\cite{imboden2014dissipation} the dissipation arising from the surrounding air is negligible for these particular measurements.\cite{kara2017generalized}
All experimentally obtained mechanical properties are shown in the Table \ref{experimental}.
\begin{table}[t]
\begin{ruledtabular}
\caption{Experimentally obtained mechanical properties of the measured devices from Ti \textit{et al.}\cite{chaoyang2021}}
\label{experimental}
\renewcommand{\arraystretch}{1.2}
\begin{tabular}{ c c c c }
$l \times w \times h$ & Mode & $f_n$ & $Q_n$ \\
($\rm \mu m^3$)   &    & (MHz) &   \\
\hline
$50 \times 0.90 \times 0.1$ &  1 & 5.2 & $3.1\times 10^4 $  \\
                               &  2 & 10.4   & $2.7\times 10^4  $ \\
                              &  3 & 15.6   & $2.6 \times 10^4 $ \\
                              &  4 & 20.8   & $ 2.9 \times 10^4 $   \\
                              \hline
$30 \times 0.90 \times 0.1$ &  1 & 8.6 & $1.1\times 10^4 $ \\
                              &  2 & 17.2  & $8.7\times 10^3  $  \\
                              &  3 & 25.9   & $7.9 \times 10^3 $  \\
                              &  4 & 34.7  & $7.1 \times 10^3 $ \\
\end{tabular}
\end{ruledtabular}
\end{table}

Fig. \ref{Figure1}(b) shows the top view of our NEMS beam geometry used in the simulations along with electrical boundary conditions for the electrothermal actuator. We also show the fixed constraints, \textit{i.e.,} hard clamp conditions for the suspended part of geometry. Fig. \ref{Figure1}(c) shows a side view of the same model geometry with a magnified image of the electrothermal actuator, where we indicate the constant temperature boundary condition. In this work, we are modeling single-sided actuation, where only one of the nanoresistors is engaged for driving the beam. The second nanoresistor is typically used for piezoresistive detection and is not of  relevance to this study other than slightly changing the geometry.  We include both nanoresistors in the simulations, as this will affect to the overall vibrational response of the beam. 

\subsection{Numerical Models}

The main workflow of the simulation process is as follows. We first model temperature oscillations within the structure due to an ac current input into the nanoresistor; here, we essentially use Ohm's Law, find the Joule heating in the nanoresistor, and then solve the heat equation with the Joule heating. This is done for various actuation frequencies between 100 kHz and 500 MHz. Next, we compute the induced stresses and bending moments using the linear thermal expansion equation based on the computed harmonically-varying temperature field. Finally, we solve the Euler-Bernoulli equation for the dynamic beam with tension to calculate the beam displacement for various excitation frequencies. All simulations are carried out within the COMSOL Multiphysics\textsuperscript{\textregistered} \cite{COMSOL} software environment using its built-in modules, including electrothermal actuation and the frequency response of NEMS.

The model geometry is directly based on the experimental devices. \cite{chaoyang2021} The linear dimensions of the doubly-clamped beam,  nanoresistors, and undercuts are identical to those in our devices. However, the larger electrodes that are connected to the nanoresistor are considerably smaller in our model than their actual size, as depicted in  Fig. \ref{Figure1}(b). The fabricated electrodes are rectangular in shape and are partially shown in Fig. \ref{Figure1}(a). We truncate the $y$ dimension of these electrodes in our simulations, because we have determined that increasing the size of the electrode beyond a certain critical value does not change the temperature field of the nanoresistor. To find the critical length of the electrode, we kept the width of the electrode at its actual value, and swept its  $y$ dimension  until the temperature field, as computed from Eq. (\ref{joule heating}) below, no longer changed. The critical length of the electrode was determined to be $ d_{c} = 6~\rm \mu m$ and is shown to scale in Fig. \ref{Figure1}(b). This computational step identifies the minimal geometry that guarantees a reliable heat transfer model. 
Table \ref{tab_devices} summarizes material properties of gold (Au) and silicon nitride (SiN) used in simulations. The Young's modulus and density of SiN, as well as the resistivity of Au for the devices were determined in previous experiments, \cite{chaoyang2021} and closely match with the values in Table \ref{tab_devices}.
\begin{table*}[t]
\caption{Material properties used in the simulations. } \label{tab_devices}
\setlength{\tabcolsep}{11pt}
\renewcommand{\arraystretch}{1.2}
\begin{ruledtabular}
\begin{tabular}{lcccc}
Property & SiN & Au & Air & Water \\
\hline
Electrical conductivity $\sigma$ (S/m) &  - & $45.6\times10^6$ & - & - \\
Thermal conductivity $k$ ($\rm W/m \cdot \rm K$) & 2.1 & 317 & 0.026 & 0.594\\
Thermal expansion coefficient $\alpha$ ($\rm K^{-1}$) & $2.3\times10^{-6}$ & $14.2\times10^{-6}$ & - & -\\
Heat capacity $c_p$ ($\rm J/kg \cdot \rm K$) & 700 & 129 & 1005 & 4200\\
Mass density $\rho$ ($\rm kg/m^3$) & 3000 & 19300 & 1.2 & 1000\\
Young's modulus $E$ (GPa) & 300 & 70 & - & -\\
Poisson's ratio $\nu$ & 0.23 & 0.44 & - & -\\
\end{tabular}
\end{ruledtabular}
\end{table*}

Prior to performing a dynamic simulation of NEMS under electrothermal actuation, we set the intrinsic stress and mode damping in the doubly-clamped beams so that the modal frequencies $f_n$ and quality factors $Q_n$ match the experimental values in Table \ref{experimental}. We determine a uniaxial tensile stress $s_{xx}^{(0)}$ in the $x$-direction normal to the $yz$ plane and a modal isotropic loss factor $\eta_n$ for the entire suspended geometry. For linear elastic materials, incorporating the isotropic dissipation into the dynamics is equivalent to making the  Young's modulus complex, \textit{i.e.,} $E(1 + i\eta_n)$, which results in a complex-valued stiffness matrix and contributes to the values of the modal quality factor $Q_n$. To determine the eigenfrequencies, we solve the following elastic equation of motion:
\begin{multline}\label{motion}
    \rho_{s} A{{{\partial ^2} W(x,t)} \over {\partial {t^2}}} - s_{xx}^{(0)}A{{{\partial ^2}W(x,t) } \over {\partial {x^2}}} \\+ E_{s}(1 + i\eta_n)I{{{\partial ^4}W(x,t) } \over {\partial {x^4}}} = {{{\partial ^2}{{\cal M}_y (x,t)}} \over {\partial {x^2}}}.
\end{multline}
Eq. (\ref{motion}) describes  the position- and time-dependent displacement, $W(x,t)$, of a rectangular beam of length $L$, cross-sectional area $A=wh$, mass density $\rho_{s}$, tension $s_{xx}^{(0)}A$ and flexural rigidity $E_{s}I$. Here, the $x$-axis is along the length of the beam and the displacement $W(x,t)$ is along the $z$-axis (Fig. \ref{Figure1}). We assume that a position-dependent harmonic bending moment ${\cal M}_y (x,t)$ is acting near the beam anchor, where the electrothermal actuation is taking place, and is responsible for driving the beam. 

In order to find the eigenfrequencies, we remove the external drive by making the right-hand side of the Eq. (\ref{motion}) zero, and implement fixed constraint boundary conditions on clamped locations of the beam, as depicted in Fig. \ref{Figure1}(b,c). This boundary condition implies $W(0,t) = W(L,t) = {{dW(0,t)}\over{dx}} = {{dW(L,t)}\over{dx}} = 0,$ where $L$ denotes the total length of the suspended part of NEMS resonator. We note that the geometry of the suspended region deviates from an ideal beam, since the cross-sectional area $A$ varies along the length $L$. As a result, FEA is particularly suitable compared to a simplified analytical model. We can express the eigenvalue solutions as $\phi_n(x)e^{-i\omega_nt}$, where $f_n = \frac{\omega_n}{2\pi}$ is the complex linear eigenfrequency, and $\phi_n(x)$ is a normalized eigenfunction along the length $L$. The modal eigenfrequencies are complex-valued due to the presence of isotropic dissipation, and we calculate the mode quality factor $Q_n$ as
\begin{equation}
    Q_n = {{\Re}(f_n)\over 2  {\Im}(f_n)},    \end{equation}
where $\Re$ and $\Im$ refer to the real and imaginary components, respectively. We conduct a parametric sweep of the uniaxial tensile stress $s_{xx}^{(0)}$ and isotropic damping factor $\eta_n$ to match the experimentally observed modal eigenfrequencies $f_n$ and quality factors $Q_n$ from Table \ref{experimental}. The intrinsic stress for a given beam determines all the modal frequencies. The damping factor, on the other hand, varies across natural modes of the resonator and is tuned for each mode. To quantify the elastic behavior of a beam with tension, we adopt the non-dimensional tension parameter $U = \frac{F_T}{EI/L^2}$ used by Barbish \textit{et al.,}\cite{Barbish2022} which represents a ratio of the tension force to the rigidity of the beam. A system with $U = 0$ is an Euler-Bernoulli beam, and a system with $U \gg 1$ behaves as a string. For our beams,  $U \sim 10^3$, and thus, for all practical purposes, they behave like strings. Our numerical model is designed to work for a broad range of $U$. Without loss of generality, we shall refer to our nanomechanical systems as NEMS beams with tension.
\begin{figure*}
    \includegraphics[width=\textwidth]{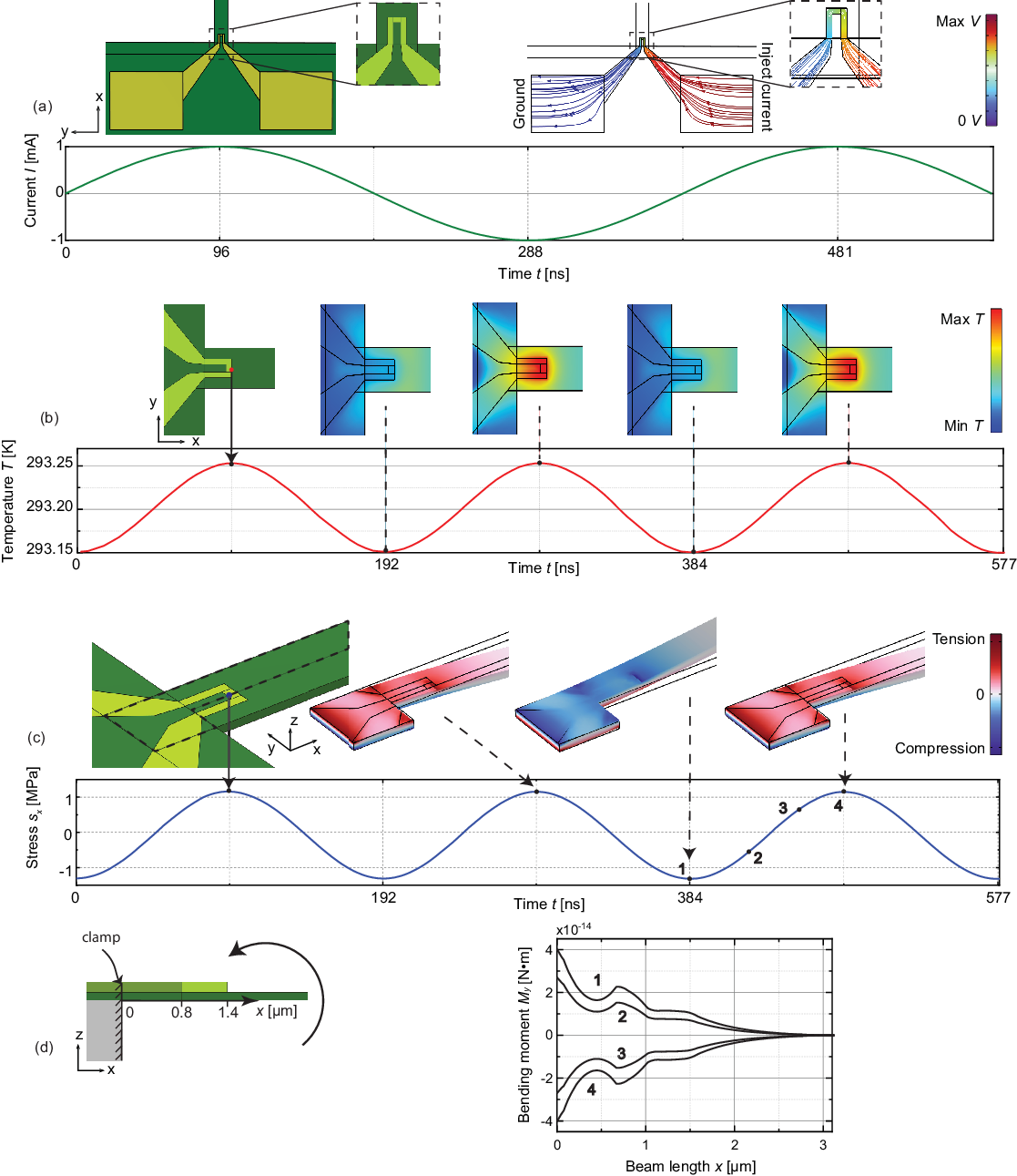}
    \caption{A synopsis of the finite-element simulation process. (a) An ac is supplied to the transducer at the frequency of $\frac{f_1}{2} = 2.6~\rm MHz$. The colormap of the current streamlines shows the electric potential across the nanoresistor. The green trace shows the change of the current amplitude in time. (b) A harmonic temperature field at frequency $f_1 = 5.2~\rm MHz$ results from the Joule heating. The red trace shows the temperature change of a point on the top of the gold actuator (marked by a red dot). Top views of the actuator's temperature field are shown at the indicated time points. (c) Thermal expansion induces harmonic mechanical stresses. The blue trace shows the stress of a point at the interface of the transducer and the beam (marked by a blue dot). Stress fields of the beam at the indicated time points are presented. Here, a positive stress value signifies tension, and a negative value implies compression. (d) We extract the bending moment along the length of the beam from the stress field. We show four instantaneous bending moment diagrams at the corresponding stress values (points 1-4).}
    \label{synopsis}
\end{figure*}

For the subsequent dynamic simulations, we demonstrate our approach on a doubly-clamped beam with tension with $l = 50~\rm \mu m$ driven exactly at its fundamental frequency. The intrinsic stress and dissipation are applied such that the fundamental frequency is at $f_1=5.2~\rm MHz$ and the mode quality factor $Q_1 = 3.1\times 10^4$, matching the experimental values in Table \ref{experimental}. A detailed synopsis of the process with intermediate results is presented in Fig. \ref{synopsis}. We begin the modeling process by introducing the following inputs for each physical domain. In the electrical domain, we inject an ac current of $I_d = 1~\rm mA$ at 2.6 MHz, which is half the resonant frequency of the fundamental mode $f_1$, into one of the electrodes, as depicted in Fig. \ref{synopsis}(a). This is equivalent to supplying a drive current of $I(t) = I_d \sin \omega_d t$, where $\omega_d = {2\pi f_d} = 2\pi ({f_1\over 2})$ is the angular drive frequency. The mechanical actuation frequency $\omega_a$ is always equal to twice the electric drive frequency $\omega_d$ due to the nature of electrothermal actuation; $\omega_n$ are the mechanical modal resonant frequencies. We calculate the instantaneous local current density within the gold layer using Ohm's law as follows:
\begin{equation}\label{current density}
    \mathbf{J}(\mathbf{r},t) = \sigma_{g} \mathbf{E}(\mathbf{r},t) + {\partial \mathbf{D}(\mathbf{r},t) \over \partial t} +  \mathbf{J_e}(\mathbf{r},t).
\end{equation}
Here, $\mathbf{J}(\mathbf{r},t)$ is the current density, $\sigma_{g}$ is the electrical conductivity of gold, $\mathbf{E}(\mathbf{r},t)$ is the electric field strength, $\mathbf{D}(\mathbf{r},t)$ is the electric flux density, and $\mathbf{J_e}(\mathbf{r},t)$ is the externally applied current density. Bold symbols here indicate vector quantities. The input current is related to $\mathbf{J_e}(\mathbf{r},t)$ as $I_d(t)= \oint \mathbf{J_e}(\mathbf{r},t)\cdot \mathbf{n} \,dS$, which indicates that $I_d(t)$ is the integral of the normal input current density integrated over the cross-sectional area of the electrode. Fig. \ref{synopsis}(a) shows the electrode on the actuation side of the beam, with the inset showing a closeup of the nanoresistor that acts as the transducer. We also compute electric current streamlines that show the current density. Here, the colormap corresponds to the electric potential. The green trace is the injected ac current in time domain with a period of ${1}/{f_d} = 384~\rm ns$.

Joule heating induces a time-dependent temperature field $T(\mathbf{r},t)$ on the entire structure that oscillates at twice the drive frequency, \textit{i.e.,} at $2f_d$. This temperature field is numerically calculated using the heat equation
\begin{equation}
\label{joule heating}
    Q_e = \rho c_p {\partial T\over\partial t} - \nabla\cdot (k\nabla T).  
\end{equation}
In Eq. (\ref{joule heating}), $Q_e(\mathbf{r},t)$ is the Joule heating rate and is defined as $Q_e(\mathbf{r},t) = \mathbf{J}(\mathbf{r},t)\cdot \mathbf{E}(\mathbf{r},t)$, where the dot indicates a scalar product;  $\rho$ is the mass density, $c_p$ is the isobaric heat capacity, and $k$ is the thermal conductivity. In addition to the Joule heating, we set a few other boundary conditions in the heat transfer domain. First, we assume that all external surfaces of the device are in thermal insulation, \textit{i.e.}, $-\textbf{n}\cdot \nabla T = 0$,  which implies that there is no heat flux going to the surrounding environment. This is a reasonable assumption for modeling electrothermal actuation in a vacuum setting. Second, we define a bath temperature, $T_b$ = 293.15 K at the bottom of the silicon nitride layer, as shown in Fig. \ref{Figure1}(c). $T_b$ is also the reference temperature, which is the initial temperature of the entire device before actuation. In other words, the structure is undeformed and free of thermal stresses at $T = T_b$.  Fig. \ref{synopsis}(b) shows the computed temperature fields $T(\mathbf{r},t)$ on the nanoresistor and the beam at various time points. The red trace shows $T(\mathbf{r},t)$ at $\mathbf{r} = \mathbf{r_1}$, which is the temperature change of the highlighted point in Fig. \ref{synopsis}(b). For $I_d = 1~\rm mA$, the harmonic temperature oscillations have a peak-to-peak amplitude of approximately $\Delta T = 0.1~\rm K$ and oscillate at the fundamental resonant frequency of $f_1 = 5.2~\rm MHz$.

Subsequently, ac-induced harmonic temperature oscillations generate thermal stresses in the actuator-beam system. This is partly due to the difference between the coefficients of thermal expansion of the gold nanoresistor and the silicon nitride beam. We switch the variable dependence from time $t$ to the actuation frequency $\omega_a$, indicating that $T(\mathbf{r},t)=\frac{\Delta T(\mathbf{r},\omega_a)}{2} \sin (\omega_at)$ due to the harmonic nature of temperature field oscillations. In the simulation software, this is equivalent to performing a frequency domain study, where the external perturbation is the harmonic thermal load that results from the computed temperature field $T (\mathbf{r},\omega_a)$. \cite{comsolsmmanual} This thermal load can be quantified by thermal stress, strain, and bending moment. We can solve for the thermal stress field by computing the strain field due to the differential thermal expansion using
\begin{equation}
\label{thermal expansion}
    \underaccent{\tilde}{\epsilon}_{lm}(\mathbf{r},\omega_a) = \underaccent{\tilde}{\alpha}_{lm} (T (\mathbf{r},\omega_a) - T_b),
\end{equation}
where $\underaccent{\tilde}{\epsilon}_{lm}$ is the second-order strain tensor, and $\underaccent{\tilde}{\alpha}_{lm}$ is the thermal expansion coefficient tensor. The thermal strain contributes to the stress field along with the initial stress $s_{xx}^{(0)}$, and the total stress can be computed as
\begin{equation}
    \underaccent{\tilde}{s}_{jk}(\mathbf{r},\omega_a) = s_{xx}^{(0)} + \underaccent{\tilde}{C}_{jklm}:\underaccent{\tilde}{\epsilon}_{lm}(\mathbf{r},\omega_a),
\end{equation}
where $\underaccent{\tilde}{s}_{jk}$ is the second-order stress tensor, $\underaccent{\tilde}{C}_{jklm}$ is the fourth-order elasticity tensor, and the colon operator signifies a double dot product; the tensor indices run from $x$ to $z$ and a sum is implied over a repeated index. We note that the thermal expansion effect is dominating in the $x$ direction due to the geometry of the structure, and thus it is relevant to show the stress field $s_x = s_{xx}(\mathbf{r},\omega_1) - s_{xx}^{(0)}$. The trace in Fig. \ref{synopsis}(c) shows the stress value for $\mathbf{r} = \mathbf{r_2},$ where $\mathbf{r_2}$ is the coordinate of the blue point. The colormaps show three instances of the harmonic stress field at the corresponding time points indicated on the trace. Here, we virtually cut the beam along the dashed lines as shown to expose the stress field inside the beam.

From computed stresses in the beam, we extract the bending moment in the $y$ direction along the length of the beam, as depicted in Fig. \ref{synopsis}(d). To this end, we  first average the $xx$-component of the stress as ${1 \over w}\int_{-w/2}^{w/2} s_{xx}(x,y,z,\omega_a)dy$ over the width of the beam $w$ to find $s_{xx}(x,z,\omega_a)$. Next, we integrate in the directions specified by the following expression:
\begin{equation}\label{bending}
    M_y(x,\omega_a)=\int_0^z (s_{xx}(x,z,\omega_a) - s_{xx}^{(0)})(z-z_n)dz.
\end{equation}
In Eq. (\ref{bending}), $z_{n}(x,\omega_a)$ is the location of the neutral axis at a given $x$, and is defined as
\begin{equation}\label{neutral}
 z_{n}(x,\omega_a) = \frac{\int_0^zs_{xx}(x,z,\omega_a)zdz}{\int_0^zs_{xx}(x,z,\omega_a)dz}.
\end{equation}
We show four instantaneous bending moment profiles in Fig. \ref{synopsis}(d) at time points 1 - 4 along the stress trace indicated in Fig. \ref{synopsis}(c). The schematic picture in Fig. \ref{synopsis}(d) shows a side view of the nanoresistor and the clamped end of the beam, which is the starting point of the bending moment diagram.

\section{Results}
\subsection{Parametric Studies}
\begin{figure}
    \includegraphics{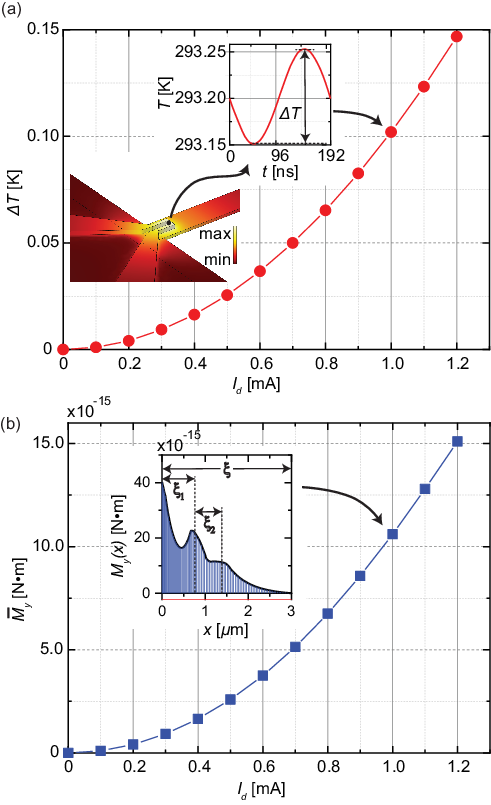}
    \caption{Change in transducer performance based on the injected ac drive amplitude $I_d$. (a) Peak-to-peak oscillatory temperature $\Delta T$ as a function of $I_d$. The lower inset shows a temperate field with the point of measurement of the steady-state temperature oscillations. The upper inset shows a snapshot of the steady-state temperature oscillations in time domain. (b) Spatially averaged bending moment $\Bar{M}_y$ as a function of $I_d$. The inset shows the peak bending moment diagram $M_y(x)$ corresponding to the drive amplitude of 1 mA.}
    \label{current}
\end{figure}
Using the approach described in the Section II, we investigate the response of the NEMS actuator as a function of drive parameters. The relevant parameters for the electrothermal actuation are the magnitude of the ac drive $I_d$ and the mechanical linear actuation frequency $f_a$. Our device for simulation is the $50$-$\rm \mu m$ beam with tension with its characteristics shown in Table \ref{experimental}. For the parametric study of the effect of $I_d$, we affix the value of $f_a = f_1$ and sweep the values of $I_d$ from zero to $1.2~\rm mA$. For the parametric study of $f_a$, we maintain a constant $I_d = 1~\rm mA$ while sweeping $f_a$ from $100~\rm kHz$ to $500~\rm MHz$. 

Fig. \ref{current}(a) shows the oscillatory peak-to-peak temperature $\Delta T$ of the actuator measured on top of the gold actuator (black point). Here, we essentially repeat the procedure outlined in Fig. \ref{synopsis} while varying the magnitude of $I_d$. The current oscillates at the frequency of 2.6 MHz and the beam resonance frequency is 5.2 MHz. This temperature $\Delta T$ is  the difference between the maximum and minimum temperature values, \textit{i.e.,} peak-to-peak amplitude, on the actuator after the system reaches steady state---as shown in the upper inset of Fig. \ref{current} (a).  The lower inset shows a snapshot of the temperature field on the actuator and the beam region for the case of $I_d = 1~\rm mA$; we note that $\Delta T\propto {I_d}^2$.

Fig. \ref{current}(b) shows the spatially averaged peak amplitude of the bending moment along the beam as a result of the oscillating temperature field. Here, we first compute the bending moment for the fundamental flexural mode at $f_1=5.2~\rm MHz$ starting from the clamped end on the actuation side using Eqs. (\ref{bending}) and (\ref{neutral}). The inset shows the peak bending moment diagram from Fig. \ref{synopsis} (d) for $I_d = 1~\rm mA$. Then, for a range of $I_d$, we compute the corresponding spatially averaged bending moment $\Bar{M_y} = \frac{\int_0^{\xi}M_y(x)dx}{\xi}$, where $\xi$ is the spatial extent of the bending moment along the $x$-axis, as depicted in the inset in Fig. \ref{current}(b). Here we also indicate the nanoresistor dimensions $\xi_1$ and $\xi_2$ on the bending moment diagram for reference. This normalization of the peak bending moment removes the position dependence. As expected, we observe that $\Bar{M_y} \propto {I_d}^2$.
\begin{figure}[t]
    \includegraphics{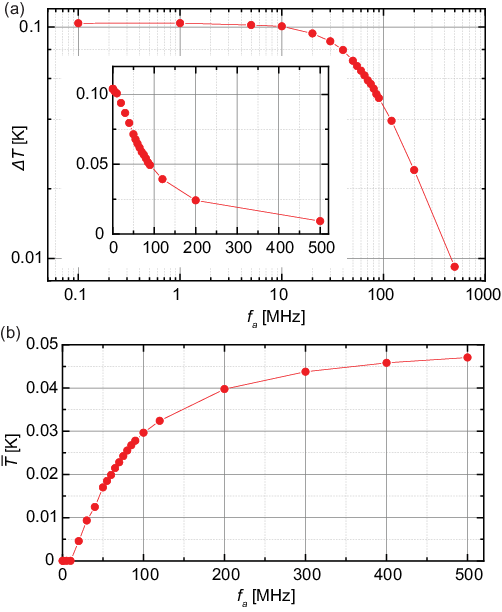}
    \caption{(a) Oscillatory temperature $\Delta T$ as a function of the actuation frequency $f_a$. The cut-off frequency is 84 MHz, and the system thermal time constant is 1.9 ns. The inset shows $\Delta T(f_a)$ in linear axes. (b) Average temperature magnitude $\Bar{T}$ as a function of the actuation frequency $f_a$.}
    \label{frequency}
\end{figure}
\begin{figure*}
    \includegraphics[width=\textwidth]{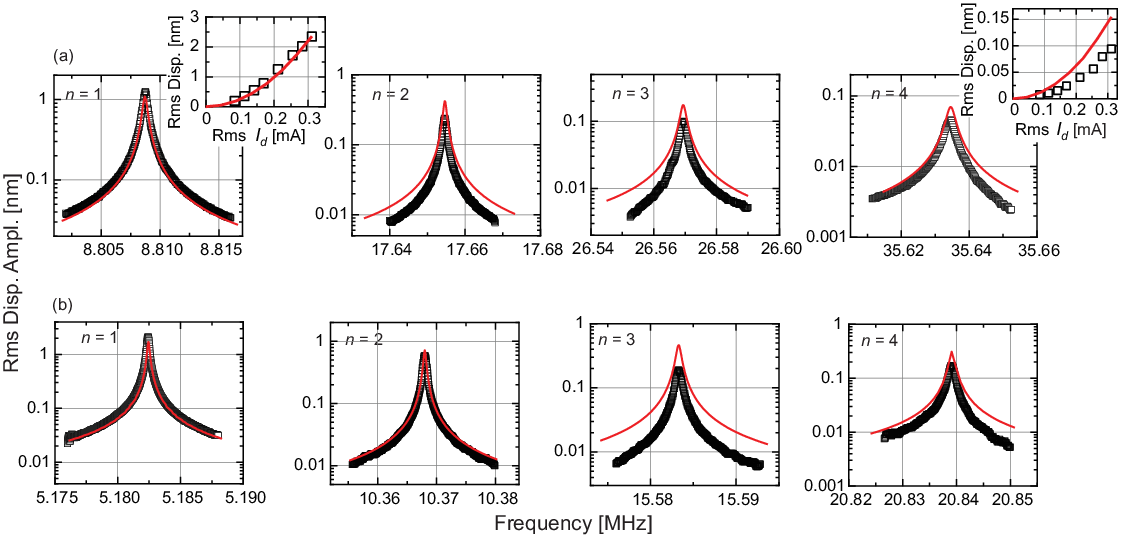}
    \caption{Comparison of experimental displacement data from Ti \textit{et al.}\cite{chaoyang2021} (black hollow squares) with FEA simulation results (red solid lines). (a) Rms displacement as a function of frequency at an antinode of the $30$-$\rm \mu m$ beam for modes 1-4. Insets show the rms displacement at the corresponding mode resonance as a function of $I_d$. (b) Rms displacement as a function of frequency at an antinode of the $50$-$\rm \mu m$ beam for modes 1-4.}
    \label{disp_ver}
\end{figure*}

Next, we investigate the thermal frequency response of electrothermal actuation for a fixed electric drive of $I_d = 1~\rm mA$. We repeat the simulation procedure for the $50$-$\rm\mu m$ beam with tension as outlined in Fig. \ref{synopsis} while varying the actuation frequency $f_a$. Here, we are interested in the peak-to-peak oscillatory temperature $\Delta T$ as a function of actuation frequency $f_a$, where $\Delta T$ is extracted from the point indicated in Fig. \ref{current}(a). The mechanical response of the device at the actuation frequency $f_a$ is linearly proportional to the thermal response as established in Fig. \ref{current}(a,b). We plot $\Delta T(f_a)$ on logarithmic axes in Fig. \ref{frequency}(a), and on linear axes in the inset, from which we extract the effective frequency range of actuation. The simulation results predict that our NEMS doubly-clamped beams with tension can be actuated efficiently up to 84 MHz (3 dB cut-off frequency), and the thermal time constant is $\tau \approx 1.9~\rm ns$. We highlight the resemblance between the nanoactuator's thermal response and the transfer function of a low-pass filter, which is in accordance with how nanoscale electrothermal actuators behave. \cite{bargatin2007efficient} 

As the oscillatory temperature range $\Delta T$ decreases at higher actuation frequencies, the dissipated power translates into a higher average temperature of the actuator, $\Bar{T}$. We formally define the average temperature as $\Bar{T} = T_p - \Delta T$, where $T_p$ is the peak temperature with respect to $T_b$. Note that the temperature oscillations cannot reach below $T_b$, thus $\Bar{T} = 0$ if $\Delta T = T_p - T_b$, as is the case shown in Fig. \ref{synopsis}(b). We show $\Bar{T}(f_a)$ on linear axes in Fig. \ref{frequency}(b).

The magnitude of $\Bar{T}$ is more substantial at higher actuation frequencies, as there is a prominent ramp-up period before the oscillatory temperature oscillations become steady. Thus, the temperature evolution in time is described by the following equation:
\begin{equation}
\label{temp_time_domain}
    T(\mathbf{r},t) = T_b + (\Bar{T} + \frac{\Delta T}{2})(1 - e^{-\frac{t}{\tau}}) + \frac{\Delta T}{2}\sin \omega_a t.
\end{equation}
The second term in Eq. (\ref{temp_time_domain}) is the dc thermal response of the system, and the time constant $\tau = 1.9~\rm ns$; the third term is the oscillatory temperature, which is responsible for electrothermal actuation.

\subsection{Validation}
We validate our FEA model by comparing beam oscillation amplitudes from experiments \cite{chaoyang2021} with simulations. Fig. \ref{disp_ver} shows a comparison between the experimental rms displacement as a function of actuation frequency of two different beams with tension with linear dimensions given in Table \ref{experimental} and our simulation results for the same beams. The transducers for these beams have the same dimensions as the representative sample in Fig. \ref{Figure1}(a). The experimental data are shown by hollow symbols, and the numerical results are shown by red solid lines. The numerical results for the beam displacement are obtained as follows. We first find all required input parameters, including the intrinsic stress, dissipation and input ac current. We then compute the harmonic temperature field from the Joule heating due to a supplied ac drive using Eqs. (\ref{current density}) and (\ref{joule heating}). Next,  we solve the Euler-Bernoulli dynamic beam equation with tension, where the external drive term is the harmonic thermal load that arises from the temperature field. \cite{comsolsmmanual} These steps are entirely carried out in the FEA software environment, and allow us to calculate the displacement field for any beam geometry, drive amplitude and actuation frequency following the procedure shown in Fig. \ref{synopsis}. In Fig. \ref{disp_ver}, we demonstrate good agreement for a $30$-$\rm \mu m$ beam and a $50$-$\rm \mu m$ beam around their first four flexural modes. We also show the rms resonance displacement amplitude as a function of the drive current $I_d$ for the first and fourth modes of the $30$-$\rm \mu m$ beam in the insets of Fig. \ref{disp_ver}(a).

\subsection{Temperature in Fluids}
\begin{figure}\label{temperature_fluid}
    \includegraphics{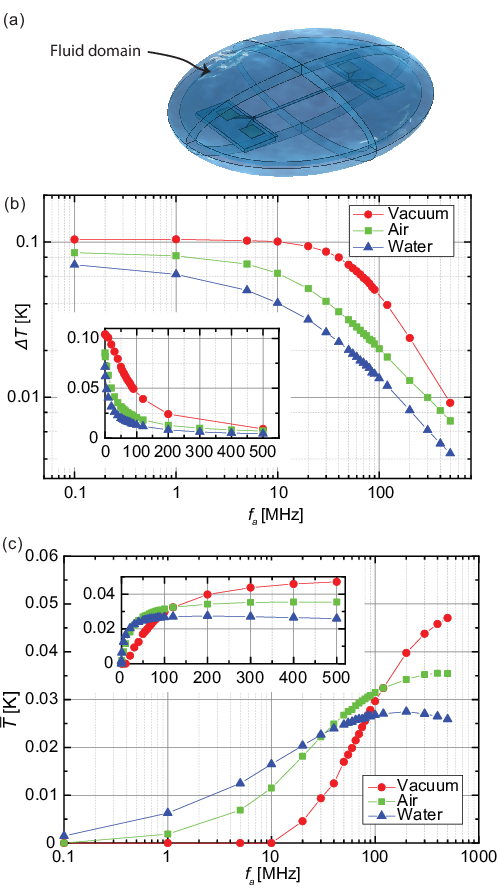}
    \caption{(a) Model geometry for the NEMS immersed in air and water. The outer shell of the indicated fluid domain is an infinite-element domain. (b) Oscillatory temperature $\Delta T$ as a function of actuation frequency $f_a$ in vacuum, air, and water. Inset shows $\Delta T(f_a)$ in linear axes. (c) Average temperature $\Bar{T}$ as a function of actuation frequency $f_a$ for vacuum, air, and water. }
    \label{fluids}
\end{figure}

We extend this analysis to an electrothermal actuator immersed in a fluid, as it is relevant for investigating nanoscale fluid-structure interactions and nanomechanical sensing applications. We investigate frequency-dependent parameters $\Delta T(f_a)$ and $\Bar{T}(f_a)$ when the NEMS structure is immersed in air and water. From the perspective of the software implementation, we surround the entire NEMS structure in an ellipsoid-shaped fluid domain and impose a uniform initial temperature $T_b$, as shown in Fig. \ref{fluids}(a). We define an infinite-element domain so that the size of the fluid volume is large compared to the nanoactuator. For computing the temperature field, we solve the heat equation (\ref{joule heating}) with additional contributions from the surrounding fluid. Relevant fluid properties used in simulations are shown in Table \ref{tab_devices}. 

We show the oscillatory temperature $\Delta T(f_a)$ on logarithmic axes in Fig. \ref{fluids}(b), and the average temperature $\Bar{T}(f_a)$ on semi-logarithmic axes in Fig. \ref{fluids}(c) for both air and water. For reference, we also plot the thermal response in vacuum. The insets in Fig. \ref{fluids}(b,c) show the the same data in linear axes. As expected, the presence of a fluid reduces the peak oscillatory temperature as well as the average temperature across the frequency range as compared to vacuum. Additionally, the fluid decreases the cut-off frequency and, as a result, increases the thermal time constant of the electrothermal actuator. 

\section{Discussion}

The developed finite-element model is broadly applicable to electrothermal transducers with linear dimensions varying from $m$ to $nm$. The innovative modeling approach here is based on a carefully-defined coupling of the thermal and mechanical domains of electrothermal actuation. More specifically, we utilize a time-dependent solver, which is the most intricate but also computationally expensive method, for finding steady-state thermal behavior. This step provides the oscillatory temperature $\Delta T$ for a given drive and frequency, and $\Delta T$ is the only input required for coupling the thermal domain to the mechanical domain. For the mechanical simulation, we incorporate the computed temperature field as a harmonic perturbation to the system, which allows us to compute the stress and displacement fields, as well as bending moments and strain fields. Beside its computational efficiency and reliability, our method for simulating electrothermal actuation overcomes a few key challenges associated with a fully-coupled transient model that is typically used in FEA software environments. For instance, it is impractical to define isotropic dissipation within COMSOL Multiphysics\textsuperscript{\textregistered} software environment, if one is working in the time domain. \cite{comsolsmmanual} This limitation will compromise the accuracy of the NEMS response as the modal quality factor $Q_n$ cannot be properly established. However, since our mechanical response is modeled in the frequency domain, this limitation does not apply. 

Furthermore, the results obtained from our simulation model can be integrated with recently developed analytical models that govern nanoscale systems. For example, for externally driven nanoscale beams with tension immersed in viscous fluids, Barbish \textit{et al.} \cite{Barbish2022} derived analytical expressions for calculating driven beam amplitude. In this analytical approach, the electrothermal drive was modeled as a uniform force per unit length $f(x)$ along the beam over a certain length, which was treated as a parameter. The authors demonstrated that such a uniform force distribution must be extended significantly further than the actuator length in order to match the experimental results, \cite{Barbish2022} which shows the complexity of modeling electrothermal drive. The FEA model developed in this manuscript provides both the force distribution and the spatial extent for various drive amplitudes and frequencies. These parameters can be computed either by numerical differentiation of the bending moment, \textit{i.e.,} $f(x) = {{d^2}{{M}_y (x)}\over {{d{x^2}}}}$ or by direct extraction of the normal force from the FEA model, and the results can be readily incorporated into analytical expressions.  Thus, informed with the output from FEA, one can better investigate the dynamic response of various nanoscale devices using analytical expressions in conjunction with numerically computed excitation parameters, such as thermal stresses, forces and bending moments.

We now provide a rough estimate for relevant physical properties related to electrothermal actuation of our NEMS devices, as computed from the FEA model. For  excitation currents $\sim 1$ mA and lower actuation frequencies below the thermal roll-off frequency, we achieve an effective temperature oscillation range of around 0.1 K without significant increase in the resonator's average temperature. This temperature increase generates a harmonic thermal stress  of $\sim 1$ MPa, and spatially-averaged bending moments of  $\sim 10^{-14}~\rm N\cdot m$. We attribute the shape of the bending moment diagram to the geometry of our NEMS doubly-clamped beams. As shown in Fig. \ref{synopsis}(d),  the undercut starts below the wider suspended SiN layer, therefore the suspended part of the structure contains both the slender SiN beam and part of the wider SiN layer. This creates an abrupt change in cross-sectional area, which results in a localized stress concentration. Additionally, the sharp end of the u-shaped actuator tip also creates local stress buildup, which  modifies the bending moment. At the cutoff frequency of $\sim$ 80 MHz, the amplitude of the temperature oscillations, stresses, and bending moments reduce by a factor of 2. The reduction in actuation efficiency is due to our system's thermal time constant of 1.9 ns, which is the time required for the temperature to equilibrate in response to ac current. At high actuation frequencies, this results in a reduced oscillatory temperature and an increased average temperature. As for the amplitude of the ac drive, we limit the excitation current such that the mechanical response of the system remains within the linear regime of operation, as nonlinear effects in NEMS are beyond the scope for this work.

Finally, we discuss the comparison of the beam displacement as a function of frequency acquired from our FEA model with the experiments. The intrinsic stress $s_{xx}^{(0)}$ and modal isotropic dissipation $\eta_n$ determine the exact resonant frequencies $\omega_n$ and quality factors $Q_n$. These parameters are tuned to match closely with the experiments, as shown in Fig. \ref{disp_ver}. The displacement amplitude is sensitive to the temperature field and depends upon thermal and electrical boundary conditions, material properties, and geometry. In addition, part of the drive signal in the experiments is reflected from the nanoresistor due to the presence of parasitic capacitance. This is not explicitly modeled in our simulations. The reflectance measurements of similar gold electrodes deposited on the same batch of NEMS devices were conducted by Ti \textit{et al.,}\cite{chaoyang2021} and the parasitic capacitance was found to be $C_{p} \approx 65~\rm pF$. We can compute the  frequency-dependent reflectance coefficient as
\begin{equation}
    \Gamma(f_d) = \frac{Z_L(f_d) - Z_0}{Z_L(f_d) + Z_0},
\end{equation}
where $Z_0 = 50~\Omega$, and $Z_L$ is the impedance of the nanoresistor. When we compute the reduction in the power dissipation at the higher modal frequency of $f_h = \frac{35.6}{2} = 17.8~\rm MHz$ compared to the lower frequency of $f_l = \frac{5.2}{2} = 2.6~\rm MHz$, we find that the power is reduced by $2\%$ at the higher frequency.  This power loss in the drive signal results in an experimentally weaker  drive and  contributes to the over-prediction of the displacement amplitude for higher frequencies (Fig. \ref{disp_ver}).
\begin{figure}[t]
    \includegraphics{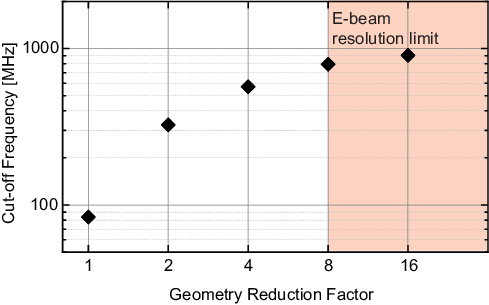}
    \caption{Geometry reduction factor vs. actuation cut-off frequency. The shaded area shows the resolution limit of modern electron beam lithography.}
    \label{size_reduction}
\end{figure}

In summary, we investigated the electrothermal actuation of NEMS and quantified the bending moments generated by the actuator and the resulting  displacements across experimentally-relevant drive amplitudes and frequencies. While the actuator can continue to excite the flexural mode of the resonators beyond 84 MHz, we show that significantly more power is required. To illustrate the effect of device scaling on the actuation bandwidth, we utilize the developed FEA model for the following parametric study. We isotropically scale down the geometry of the NEMS device by a factor of 2 and compute the corresponding actuation cut-off frequency following the modeling procedure described in this manuscript. Fig. \ref{size_reduction} shows the cut-off frequency of electrothermal actuation as a function of the geometry reduction factor plotted using logarithmic axes. There is a practical limit for miniaturization due to the resolution of electron beam lithography of $\sim$ 15 nm.\cite{fomenkov2017euv} Thus, scaling the width of the nanoresistor beyond this is unrealistic.  We show this limit in Fig. \ref{size_reduction} by the shaded region. As electromechanical systems are further miniaturized with higher operation frequencies, our approach will allow for designing more efficient electrothermal actuators and devices.

\begin{acknowledgments}
We acknowledge support from the US NSF (CMMI-2001403 and CMMI-1934271). M. M. acknowledges support from Boston University Nanotechnology Innovation Center BUnano Cross-Disciplinary Fellowship. We thank S. Kyoung for preliminary work on the problem and H. Gress for the SEM image of the representative NEMS device.
\end{acknowledgments}

\section*{author declarations}
The authors have no conflict to disclose.

\section*{Appendix: Modeling Procedure}
This section is intended to provide a more concrete set of instructions on how to build the FEA model described in this manuscript.

The FEA model contains three distinct solvers: i) a prestressed eigenfrequency solver for determining intrinsic stress and dissipation, ii) a time-dependent solver for determining the temperature field of the system via Joule heating, and iii) a frequency-domain solver that relates the harmonic thermal load to the stress field and displacement field. The main text contains all necessary details for i), and we thus focus our instructions for parts ii) and iii).

The time-dependent Joule heating solver combines the \textit{electrical currents} and \textit{heat transfer in solids} interfaces, and solves for the temperature field for a given drive amplitude $I_d$ and frequency $f_d$ using the Eqs. \ref{current density} and \ref{joule heating}. The solution time span is set to $\frac{40}{f_d}$ so that the system can reach its steady temperature oscillations with a peak-to-peak amplitude of $\Delta T$.

We extract the solution of the oscillatory temperature field from ii) by using the \textit{withsol} command from a relevant time domain simulation; then, we provide the extracted thermal field as an input into the \textit{thermal expansion} node of the \textit{solid mechanics} module, which can be efficiently solved in the frequency domain by COMSOL. Here, the oscillatory temperature field will contribute to the external load of the system and act as the excitation source, and thus one can compute the mechanical response of the system by solving the Eq. \ref{motion} at various actuation frequencies $\omega_a$, including the modal frequencies $\omega_n$. This step gives the solution of the stress and displacement fields.

\bibliographystyle{apsrev4-2}
\bibliography{Reference.bib}

\begin{thebibliography}{72}%
\makeatletter
\providecommand \@ifxundefined [1]{%
 \@ifx{#1\undefined}
}%
\providecommand \@ifnum [1]{%
 \ifnum #1\expandafter \@firstoftwo
 \else \expandafter \@secondoftwo
 \fi
}%
\providecommand \@ifx [1]{%
 \ifx #1\expandafter \@firstoftwo
 \else \expandafter \@secondoftwo
 \fi
}%
\providecommand \natexlab [1]{#1}%
\providecommand \enquote  [1]{``#1''}%
\providecommand \bibnamefont  [1]{#1}%
\providecommand \bibfnamefont [1]{#1}%
\providecommand \citenamefont [1]{#1}%
\providecommand \href@noop [0]{\@secondoftwo}%
\providecommand \href [0]{\begingroup \@sanitize@url \@href}%
\providecommand \@href[1]{\@@startlink{#1}\@@href}%
\providecommand \@@href[1]{\endgroup#1\@@endlink}%
\providecommand \@sanitize@url [0]{\catcode `\\12\catcode `\$12\catcode `\&12\catcode `\#12\catcode `\^12\catcode `\_12\catcode `\%12\relax}%
\providecommand \@@startlink[1]{}%
\providecommand \@@endlink[0]{}%
\providecommand \url  [0]{\begingroup\@sanitize@url \@url }%
\providecommand \@url [1]{\endgroup\@href {#1}{\urlprefix }}%
\providecommand \urlprefix  [0]{URL }%
\providecommand \Eprint [0]{\href }%
\providecommand \doibase [0]{https://doi.org/}%
\providecommand \selectlanguage [0]{\@gobble}%
\providecommand \bibinfo  [0]{\@secondoftwo}%
\providecommand \bibfield  [0]{\@secondoftwo}%
\providecommand \translation [1]{[#1]}%
\providecommand \BibitemOpen [0]{}%
\providecommand \bibitemStop [0]{}%
\providecommand \bibitemNoStop [0]{.\EOS\space}%
\providecommand \EOS [0]{\spacefactor3000\relax}%
\providecommand \BibitemShut  [1]{\csname bibitem#1\endcsname}%
\let\auto@bib@innerbib\@empty
%</preamble>
\bibitem [{\citenamefont {Kouh}\ \emph {et~al.}(2017)\citenamefont {Kouh}, \citenamefont {Hanay},\ and\ \citenamefont {Ekinci}}]{kouh2017nanomechanical}%
  \BibitemOpen
  \bibfield  {author} {\bibinfo {author} {\bibfnamefont {T.}~\bibnamefont {Kouh}}, \bibinfo {author} {\bibfnamefont {M.~S.}\ \bibnamefont {Hanay}},\ and\ \bibinfo {author} {\bibfnamefont {K.~L.}\ \bibnamefont {Ekinci}},\ }\href@noop {} {\bibfield  {journal} {\bibinfo  {journal} {Micromachines}\ }\textbf {\bibinfo {volume} {8}},\ \bibinfo {pages} {108} (\bibinfo {year} {2017})}\BibitemShut {NoStop}%
\bibitem [{\citenamefont {Ekinci}(2005)}]{Ekinci2005small}%
  \BibitemOpen
  \bibfield  {author} {\bibinfo {author} {\bibfnamefont {K.}~\bibnamefont {Ekinci}},\ }\href@noop {} {\bibfield  {journal} {\bibinfo  {journal} {Small}\ }\textbf {\bibinfo {volume} {1}},\ \bibinfo {pages} {786} (\bibinfo {year} {2005})}\BibitemShut {NoStop}%
\bibitem [{\citenamefont {Liang}\ \emph {et~al.}(2012)\citenamefont {Liang}, \citenamefont {Huang}, \citenamefont {Li}, \citenamefont {Huang}, \citenamefont {Wu}, \citenamefont {Fang}, \citenamefont {Oh}, \citenamefont {Kozlov}, \citenamefont {Ma}, \citenamefont {Li} \emph {et~al.}}]{Liang2012}%
  \BibitemOpen
  \bibfield  {author} {\bibinfo {author} {\bibfnamefont {J.}~\bibnamefont {Liang}}, \bibinfo {author} {\bibfnamefont {L.}~\bibnamefont {Huang}}, \bibinfo {author} {\bibfnamefont {N.}~\bibnamefont {Li}}, \bibinfo {author} {\bibfnamefont {Y.}~\bibnamefont {Huang}}, \bibinfo {author} {\bibfnamefont {Y.}~\bibnamefont {Wu}}, \bibinfo {author} {\bibfnamefont {S.}~\bibnamefont {Fang}}, \bibinfo {author} {\bibfnamefont {J.}~\bibnamefont {Oh}}, \bibinfo {author} {\bibfnamefont {M.}~\bibnamefont {Kozlov}}, \bibinfo {author} {\bibfnamefont {Y.}~\bibnamefont {Ma}}, \bibinfo {author} {\bibfnamefont {F.}~\bibnamefont {Li}}, \emph {et~al.},\ }\href@noop {} {\bibfield  {journal} {\bibinfo  {journal} {ACS Nano}\ }\textbf {\bibinfo {volume} {6}},\ \bibinfo {pages} {4508} (\bibinfo {year} {2012})}\BibitemShut {NoStop}%
\bibitem [{\citenamefont {Jin}\ \emph {et~al.}(2020)\citenamefont {Jin}, \citenamefont {Zhang}, \citenamefont {Xu}, \citenamefont {Trase}, \citenamefont {Huang}, \citenamefont {Dong}, \citenamefont {Liu}, \citenamefont {Usherwood}, \citenamefont {Zhang},\ and\ \citenamefont {Chen}}]{Jin2020}%
  \BibitemOpen
  \bibfield  {author} {\bibinfo {author} {\bibfnamefont {C.}~\bibnamefont {Jin}}, \bibinfo {author} {\bibfnamefont {J.}~\bibnamefont {Zhang}}, \bibinfo {author} {\bibfnamefont {Z.}~\bibnamefont {Xu}}, \bibinfo {author} {\bibfnamefont {I.}~\bibnamefont {Trase}}, \bibinfo {author} {\bibfnamefont {S.}~\bibnamefont {Huang}}, \bibinfo {author} {\bibfnamefont {L.}~\bibnamefont {Dong}}, \bibinfo {author} {\bibfnamefont {Z.}~\bibnamefont {Liu}}, \bibinfo {author} {\bibfnamefont {S.~E.}\ \bibnamefont {Usherwood}}, \bibinfo {author} {\bibfnamefont {J.~X.~J.}\ \bibnamefont {Zhang}},\ and\ \bibinfo {author} {\bibfnamefont {Z.}~\bibnamefont {Chen}},\ }\href {https://doi.org/10.1002/aisy.201900162} {\bibfield  {journal} {\bibinfo  {journal} {Adv. Intell. Syst.}\ }\textbf {\bibinfo {volume} {2}},\ \bibinfo {pages} {1900162} (\bibinfo {year} {2020})}\BibitemShut {NoStop}%
\bibitem [{\citenamefont {Pop}\ \emph {et~al.}(2022)\citenamefont {Pop}, \citenamefont {Herrera},\ and\ \citenamefont {Rinaldi}}]{pop2022lithium}%
  \BibitemOpen
  \bibfield  {author} {\bibinfo {author} {\bibfnamefont {F.}~\bibnamefont {Pop}}, \bibinfo {author} {\bibfnamefont {B.}~\bibnamefont {Herrera}},\ and\ \bibinfo {author} {\bibfnamefont {M.}~\bibnamefont {Rinaldi}},\ }\href@noop {} {\bibfield  {journal} {\bibinfo  {journal} {Nat. Commun.}\ }\textbf {\bibinfo {volume} {13}},\ \bibinfo {pages} {1782} (\bibinfo {year} {2022})}\BibitemShut {NoStop}%
\bibitem [{\citenamefont {Mahboob}\ and\ \citenamefont {Yamaguchi}(2008)}]{Mahboob2008}%
  \BibitemOpen
  \bibfield  {author} {\bibinfo {author} {\bibfnamefont {I.}~\bibnamefont {Mahboob}}\ and\ \bibinfo {author} {\bibfnamefont {H.}~\bibnamefont {Yamaguchi}},\ }\href {https://doi.org/10.1038/nnano.2008.84} {\bibfield  {journal} {\bibinfo  {journal} {Nat. Nanotechnol.}\ }\textbf {\bibinfo {volume} {3}},\ \bibinfo {pages} {275} (\bibinfo {year} {2008})}\BibitemShut {NoStop}%
\bibitem [{\citenamefont {Qian}\ \emph {et~al.}(2017)\citenamefont {Qian}, \citenamefont {Peschot}, \citenamefont {Osoba}, \citenamefont {Ye},\ and\ \citenamefont {Liu}}]{Qian2017}%
  \BibitemOpen
  \bibfield  {author} {\bibinfo {author} {\bibfnamefont {C.}~\bibnamefont {Qian}}, \bibinfo {author} {\bibfnamefont {A.}~\bibnamefont {Peschot}}, \bibinfo {author} {\bibfnamefont {B.}~\bibnamefont {Osoba}}, \bibinfo {author} {\bibfnamefont {Z.~A.}\ \bibnamefont {Ye}},\ and\ \bibinfo {author} {\bibfnamefont {T.~J.~K.}\ \bibnamefont {Liu}},\ }\href {https://doi.org/10.1109/TED.2017.2657554} {\bibfield  {journal} {\bibinfo  {journal} {IEEE Trans. Electron Devices}\ }\textbf {\bibinfo {volume} {64}},\ \bibinfo {pages} {1323} (\bibinfo {year} {2017})}\BibitemShut {NoStop}%
\bibitem [{\citenamefont {Maluf}\ and\ \citenamefont {Williams}(2004)}]{maluf2004introduction}%
  \BibitemOpen
  \bibfield  {author} {\bibinfo {author} {\bibfnamefont {N.}~\bibnamefont {Maluf}}\ and\ \bibinfo {author} {\bibfnamefont {K.}~\bibnamefont {Williams}},\ }\href@noop {} {\emph {\bibinfo {title} {Introduction to microelectromechanical systems engineering}}}\ (\bibinfo  {publisher} {Artech House},\ \bibinfo {year} {2004})\BibitemShut {NoStop}%
\bibitem [{\citenamefont {Villanueva}\ \emph {et~al.}(2008)\citenamefont {Villanueva}, \citenamefont {Plaza}, \citenamefont {Montserrat}, \citenamefont {Perez-Murano},\ and\ \citenamefont {Bausells}}]{villanueva2008crystalline}%
  \BibitemOpen
  \bibfield  {author} {\bibinfo {author} {\bibfnamefont {G.}~\bibnamefont {Villanueva}}, \bibinfo {author} {\bibfnamefont {J.}~\bibnamefont {Plaza}}, \bibinfo {author} {\bibfnamefont {J.}~\bibnamefont {Montserrat}}, \bibinfo {author} {\bibfnamefont {F.}~\bibnamefont {Perez-Murano}},\ and\ \bibinfo {author} {\bibfnamefont {J.}~\bibnamefont {Bausells}},\ }\href@noop {} {\bibfield  {journal} {\bibinfo  {journal} {Microelectron. Eng.}\ }\textbf {\bibinfo {volume} {85}},\ \bibinfo {pages} {1120} (\bibinfo {year} {2008})}\BibitemShut {NoStop}%
\bibitem [{\citenamefont {Chui}\ \emph {et~al.}(1998)\citenamefont {Chui}, \citenamefont {Kenny}, \citenamefont {Mamin}, \citenamefont {Terris},\ and\ \citenamefont {Rugar}}]{chui1998independent}%
  \BibitemOpen
  \bibfield  {author} {\bibinfo {author} {\bibfnamefont {B.}~\bibnamefont {Chui}}, \bibinfo {author} {\bibfnamefont {T.}~\bibnamefont {Kenny}}, \bibinfo {author} {\bibfnamefont {H.}~\bibnamefont {Mamin}}, \bibinfo {author} {\bibfnamefont {B.}~\bibnamefont {Terris}},\ and\ \bibinfo {author} {\bibfnamefont {D.}~\bibnamefont {Rugar}},\ }\href@noop {} {\bibfield  {journal} {\bibinfo  {journal} {Appl. Phys. Lett.}\ }\textbf {\bibinfo {volume} {72}},\ \bibinfo {pages} {1388} (\bibinfo {year} {1998})}\BibitemShut {NoStop}%
\bibitem [{\citenamefont {Mile}\ \emph {et~al.}(2010)\citenamefont {Mile}, \citenamefont {Jourdan}, \citenamefont {Bargatin}, \citenamefont {Labarthe}, \citenamefont {Marcoux}, \citenamefont {Andreucci}, \citenamefont {Hentz}, \citenamefont {Kharrat}, \citenamefont {Colinet},\ and\ \citenamefont {Duraffourg}}]{mile2010plane}%
  \BibitemOpen
  \bibfield  {author} {\bibinfo {author} {\bibfnamefont {E.}~\bibnamefont {Mile}}, \bibinfo {author} {\bibfnamefont {G.}~\bibnamefont {Jourdan}}, \bibinfo {author} {\bibfnamefont {I.}~\bibnamefont {Bargatin}}, \bibinfo {author} {\bibfnamefont {S.}~\bibnamefont {Labarthe}}, \bibinfo {author} {\bibfnamefont {C.}~\bibnamefont {Marcoux}}, \bibinfo {author} {\bibfnamefont {P.}~\bibnamefont {Andreucci}}, \bibinfo {author} {\bibfnamefont {S.}~\bibnamefont {Hentz}}, \bibinfo {author} {\bibfnamefont {C.}~\bibnamefont {Kharrat}}, \bibinfo {author} {\bibfnamefont {E.}~\bibnamefont {Colinet}},\ and\ \bibinfo {author} {\bibfnamefont {L.}~\bibnamefont {Duraffourg}},\ }\href@noop {} {\bibfield  {journal} {\bibinfo  {journal} {Nanotechnology}\ }\textbf {\bibinfo {volume} {21}},\ \bibinfo {pages} {165504} (\bibinfo {year} {2010})}\BibitemShut {NoStop}%
\bibitem [{\citenamefont {Li}\ \emph {et~al.}(2007)\citenamefont {Li}, \citenamefont {Tang},\ and\ \citenamefont {Roukes}}]{li2007ultra}%
  \BibitemOpen
  \bibfield  {author} {\bibinfo {author} {\bibfnamefont {M.}~\bibnamefont {Li}}, \bibinfo {author} {\bibfnamefont {H.~X.}\ \bibnamefont {Tang}},\ and\ \bibinfo {author} {\bibfnamefont {M.~L.}\ \bibnamefont {Roukes}},\ }\href@noop {} {\bibfield  {journal} {\bibinfo  {journal} {Nat. Nanotechnol.}\ }\textbf {\bibinfo {volume} {2}},\ \bibinfo {pages} {114} (\bibinfo {year} {2007})}\BibitemShut {NoStop}%
\bibitem [{\citenamefont {Bargatin}\ \emph {et~al.}(2007)\citenamefont {Bargatin}, \citenamefont {Kozinsky},\ and\ \citenamefont {Roukes}}]{bargatin2007efficient}%
  \BibitemOpen
  \bibfield  {author} {\bibinfo {author} {\bibfnamefont {I.}~\bibnamefont {Bargatin}}, \bibinfo {author} {\bibfnamefont {I.}~\bibnamefont {Kozinsky}},\ and\ \bibinfo {author} {\bibfnamefont {M.}~\bibnamefont {Roukes}},\ }\href@noop {} {\bibfield  {journal} {\bibinfo  {journal} {Appl. Phys. Lett.}\ }\textbf {\bibinfo {volume} {90}},\ \bibinfo {pages} {093116} (\bibinfo {year} {2007})}\BibitemShut {NoStop}%
\bibitem [{\citenamefont {Tang}\ \emph {et~al.}(2009)\citenamefont {Tang}, \citenamefont {Li},\ and\ \citenamefont {Roukes}}]{tang2009metallic}%
  \BibitemOpen
  \bibfield  {author} {\bibinfo {author} {\bibfnamefont {H.}~\bibnamefont {Tang}}, \bibinfo {author} {\bibfnamefont {M.}~\bibnamefont {Li}},\ and\ \bibinfo {author} {\bibfnamefont {M.~L.}\ \bibnamefont {Roukes}},\ }\href@noop {} {\bibinfo {title} {Metallic thin film piezoresistive transduction in micromechanical and nanomechanical devices and its application in self-sensing spm probes}} (\bibinfo {year} {2009}),\ \bibinfo {note} {uS Patent 7,617,736}\BibitemShut {NoStop}%
\bibitem [{\citenamefont {Nan}\ \emph {et~al.}(2013)\citenamefont {Nan}, \citenamefont {Hui}, \citenamefont {Rinaldi},\ and\ \citenamefont {Sun}}]{Nan2013}%
  \BibitemOpen
  \bibfield  {author} {\bibinfo {author} {\bibfnamefont {T.}~\bibnamefont {Nan}}, \bibinfo {author} {\bibfnamefont {Y.}~\bibnamefont {Hui}}, \bibinfo {author} {\bibfnamefont {M.}~\bibnamefont {Rinaldi}},\ and\ \bibinfo {author} {\bibfnamefont {N.~X.}\ \bibnamefont {Sun}},\ }\href@noop {} {\bibfield  {journal} {\bibinfo  {journal} {Sci. Rep.}\ }\textbf {\bibinfo {volume} {3}},\ \bibinfo {pages} {1985} (\bibinfo {year} {2013})}\BibitemShut {NoStop}%
\bibitem [{\citenamefont {Huang}\ \emph {et~al.}(2005)\citenamefont {Huang}, \citenamefont {Feng}, \citenamefont {Zorman}, \citenamefont {Mehregany},\ and\ \citenamefont {Roukes}}]{Huang2005}%
  \BibitemOpen
  \bibfield  {author} {\bibinfo {author} {\bibfnamefont {X.}~\bibnamefont {Huang}}, \bibinfo {author} {\bibfnamefont {X.}~\bibnamefont {Feng}}, \bibinfo {author} {\bibfnamefont {C.}~\bibnamefont {Zorman}}, \bibinfo {author} {\bibfnamefont {M.}~\bibnamefont {Mehregany}},\ and\ \bibinfo {author} {\bibfnamefont {M.}~\bibnamefont {Roukes}},\ }\href@noop {} {\bibfield  {journal} {\bibinfo  {journal} {New J. Phys.}\ }\textbf {\bibinfo {volume} {7}},\ \bibinfo {pages} {247} (\bibinfo {year} {2005})}\BibitemShut {NoStop}%
\bibitem [{\citenamefont {Unterreithmeier}\ \emph {et~al.}(2009)\citenamefont {Unterreithmeier}, \citenamefont {Weig},\ and\ \citenamefont {Kotthaus}}]{Unterreithmeier2009}%
  \BibitemOpen
  \bibfield  {author} {\bibinfo {author} {\bibfnamefont {Q.~P.}\ \bibnamefont {Unterreithmeier}}, \bibinfo {author} {\bibfnamefont {E.~M.}\ \bibnamefont {Weig}},\ and\ \bibinfo {author} {\bibfnamefont {J.~P.}\ \bibnamefont {Kotthaus}},\ }\href {https://doi.org/10.1038/nature07932} {\bibfield  {journal} {\bibinfo  {journal} {Nature}\ }\textbf {\bibinfo {volume} {458}},\ \bibinfo {pages} {1001} (\bibinfo {year} {2009})}\BibitemShut {NoStop}%
\bibitem [{\citenamefont {Feng}\ \emph {et~al.}(2007)\citenamefont {Feng}, \citenamefont {He}, \citenamefont {Yang},\ and\ \citenamefont {Roukes}}]{Feng2007}%
  \BibitemOpen
  \bibfield  {author} {\bibinfo {author} {\bibfnamefont {X.~L.}\ \bibnamefont {Feng}}, \bibinfo {author} {\bibfnamefont {R.}~\bibnamefont {He}}, \bibinfo {author} {\bibfnamefont {P.}~\bibnamefont {Yang}},\ and\ \bibinfo {author} {\bibfnamefont {M.~L.}\ \bibnamefont {Roukes}},\ }\href {https://doi.org/10.1021/nl0706695} {\bibfield  {journal} {\bibinfo  {journal} {Nano Lett.}\ }\textbf {\bibinfo {volume} {7}},\ \bibinfo {pages} {1953} (\bibinfo {year} {2007})}\BibitemShut {NoStop}%
\bibitem [{\citenamefont {Masmanidis}\ \emph {et~al.}(2005)\citenamefont {Masmanidis}, \citenamefont {Tang}, \citenamefont {Myers}, \citenamefont {Li}, \citenamefont {De~Greve}, \citenamefont {Vermeulen}, \citenamefont {Van~Roy},\ and\ \citenamefont {Roukes}}]{Masmanidis2005}%
  \BibitemOpen
  \bibfield  {author} {\bibinfo {author} {\bibfnamefont {S.}~\bibnamefont {Masmanidis}}, \bibinfo {author} {\bibfnamefont {H.}~\bibnamefont {Tang}}, \bibinfo {author} {\bibfnamefont {E.}~\bibnamefont {Myers}}, \bibinfo {author} {\bibfnamefont {M.}~\bibnamefont {Li}}, \bibinfo {author} {\bibfnamefont {K.}~\bibnamefont {De~Greve}}, \bibinfo {author} {\bibfnamefont {G.}~\bibnamefont {Vermeulen}}, \bibinfo {author} {\bibfnamefont {W.}~\bibnamefont {Van~Roy}},\ and\ \bibinfo {author} {\bibfnamefont {M.}~\bibnamefont {Roukes}},\ }\href@noop {} {\bibfield  {journal} {\bibinfo  {journal} {Phys. Rev. Lett.}\ }\textbf {\bibinfo {volume} {95}},\ \bibinfo {pages} {187206} (\bibinfo {year} {2005})}\BibitemShut {NoStop}%
\bibitem [{\citenamefont {Li}\ \emph {et~al.}(2017)\citenamefont {Li}, \citenamefont {Matyushov}, \citenamefont {Dong}, \citenamefont {Chen}, \citenamefont {Lin}, \citenamefont {Nan}, \citenamefont {Qian}, \citenamefont {Rinaldi}, \citenamefont {Lin},\ and\ \citenamefont {Sun}}]{Li2017}%
  \BibitemOpen
  \bibfield  {author} {\bibinfo {author} {\bibfnamefont {M.}~\bibnamefont {Li}}, \bibinfo {author} {\bibfnamefont {A.}~\bibnamefont {Matyushov}}, \bibinfo {author} {\bibfnamefont {C.}~\bibnamefont {Dong}}, \bibinfo {author} {\bibfnamefont {H.}~\bibnamefont {Chen}}, \bibinfo {author} {\bibfnamefont {H.}~\bibnamefont {Lin}}, \bibinfo {author} {\bibfnamefont {T.}~\bibnamefont {Nan}}, \bibinfo {author} {\bibfnamefont {Z.}~\bibnamefont {Qian}}, \bibinfo {author} {\bibfnamefont {M.}~\bibnamefont {Rinaldi}}, \bibinfo {author} {\bibfnamefont {Y.}~\bibnamefont {Lin}},\ and\ \bibinfo {author} {\bibfnamefont {N.~X.}\ \bibnamefont {Sun}},\ }\href@noop {} {\bibfield  {journal} {\bibinfo  {journal} {Appl. Phys. Lett.}\ }\textbf {\bibinfo {volume} {110}},\ \bibinfo {pages} {143510} (\bibinfo {year} {2017})}\BibitemShut {NoStop}%
\bibitem [{\citenamefont {Andreassen}\ and\ \citenamefont {Mielnik}(2014)}]{Andreassen2014}%
  \BibitemOpen
  \bibfield  {author} {\bibinfo {author} {\bibfnamefont {E.}~\bibnamefont {Andreassen}}\ and\ \bibinfo {author} {\bibfnamefont {M.~M.}\ \bibnamefont {Mielnik}},\ }in\ \href@noop {} {\emph {\bibinfo {booktitle} {Proceedings of the 5th Electronics System-integration Technology Conference (ESTC)}}}\ (\bibinfo {organization} {IEEE},\ \bibinfo {year} {2014})\ pp.\ \bibinfo {pages} {1--5}\BibitemShut {NoStop}%
\bibitem [{\citenamefont {Ti}\ \emph {et~al.}(2022)\citenamefont {Ti}, \citenamefont {McDaniel}, \citenamefont {Liem}, \citenamefont {Gress}, \citenamefont {Ma}, \citenamefont {Kyoung}, \citenamefont {Svitelskiy}, \citenamefont {Yanik}, \citenamefont {Kaya}, \citenamefont {Hanay} \emph {et~al.}}]{ti2022}%
  \BibitemOpen
  \bibfield  {author} {\bibinfo {author} {\bibfnamefont {C.}~\bibnamefont {Ti}}, \bibinfo {author} {\bibfnamefont {J.}~\bibnamefont {McDaniel}}, \bibinfo {author} {\bibfnamefont {A.}~\bibnamefont {Liem}}, \bibinfo {author} {\bibfnamefont {H.}~\bibnamefont {Gress}}, \bibinfo {author} {\bibfnamefont {M.}~\bibnamefont {Ma}}, \bibinfo {author} {\bibfnamefont {S.}~\bibnamefont {Kyoung}}, \bibinfo {author} {\bibfnamefont {O.}~\bibnamefont {Svitelskiy}}, \bibinfo {author} {\bibfnamefont {C.}~\bibnamefont {Yanik}}, \bibinfo {author} {\bibfnamefont {I.}~\bibnamefont {Kaya}}, \bibinfo {author} {\bibfnamefont {M.}~\bibnamefont {Hanay}}, \emph {et~al.},\ }\href@noop {} {\bibfield  {journal} {\bibinfo  {journal} {Appl. Phys. Lett.}\ }\textbf {\bibinfo {volume} {121}},\ \bibinfo {pages} {023506} (\bibinfo {year} {2022})}\BibitemShut {NoStop}%
\bibitem [{\citenamefont {Ari}\ \emph {et~al.}(2020)\citenamefont {Ari}, \citenamefont {Hanay}, \citenamefont {Paul},\ and\ \citenamefont {Ekinci}}]{ari2020nanomechanical}%
  \BibitemOpen
  \bibfield  {author} {\bibinfo {author} {\bibfnamefont {A.~B.}\ \bibnamefont {Ari}}, \bibinfo {author} {\bibfnamefont {M.~S.}\ \bibnamefont {Hanay}}, \bibinfo {author} {\bibfnamefont {M.~R.}\ \bibnamefont {Paul}},\ and\ \bibinfo {author} {\bibfnamefont {K.~L.}\ \bibnamefont {Ekinci}},\ }\href@noop {} {\bibfield  {journal} {\bibinfo  {journal} {Nano Lett.}\ }\textbf {\bibinfo {volume} {21}},\ \bibinfo {pages} {375} (\bibinfo {year} {2020})}\BibitemShut {NoStop}%
\bibitem [{\citenamefont {Singh}\ \emph {et~al.}(2014)\citenamefont {Singh}, \citenamefont {Chua}, \citenamefont {Liang}, \citenamefont {Jayaraman}, \citenamefont {Do},\ and\ \citenamefont {Kim}}]{Singh2014}%
  \BibitemOpen
  \bibfield  {author} {\bibinfo {author} {\bibfnamefont {P.}~\bibnamefont {Singh}}, \bibinfo {author} {\bibfnamefont {G.~L.}\ \bibnamefont {Chua}}, \bibinfo {author} {\bibfnamefont {Y.~S.}\ \bibnamefont {Liang}}, \bibinfo {author} {\bibfnamefont {K.~G.}\ \bibnamefont {Jayaraman}}, \bibinfo {author} {\bibfnamefont {A.~T.}\ \bibnamefont {Do}},\ and\ \bibinfo {author} {\bibfnamefont {T.~T.-H.}\ \bibnamefont {Kim}},\ }\href@noop {} {\bibfield  {journal} {\bibinfo  {journal} {J. Micromech. Microeng.}\ }\textbf {\bibinfo {volume} {24}},\ \bibinfo {pages} {115007} (\bibinfo {year} {2014})}\BibitemShut {NoStop}%
\bibitem [{\citenamefont {Rajaram}\ \emph {et~al.}(2022)\citenamefont {Rajaram}, \citenamefont {Kang}, \citenamefont {Calisgan}, \citenamefont {Risso}, \citenamefont {Qian},\ and\ \citenamefont {Rinaldi}}]{Rajaram2022}%
  \BibitemOpen
  \bibfield  {author} {\bibinfo {author} {\bibfnamefont {V.}~\bibnamefont {Rajaram}}, \bibinfo {author} {\bibfnamefont {S.}~\bibnamefont {Kang}}, \bibinfo {author} {\bibfnamefont {S.~D.}\ \bibnamefont {Calisgan}}, \bibinfo {author} {\bibfnamefont {A.}~\bibnamefont {Risso}}, \bibinfo {author} {\bibfnamefont {Z.}~\bibnamefont {Qian}},\ and\ \bibinfo {author} {\bibfnamefont {M.}~\bibnamefont {Rinaldi}},\ }\href@noop {} {\bibfield  {journal} {\bibinfo  {journal} {J. Appl. Phys.}\ }\textbf {\bibinfo {volume} {131}},\ \bibinfo {pages} {244501} (\bibinfo {year} {2022})}\BibitemShut {NoStop}%
\bibitem [{\citenamefont {Koochi}\ \emph {et~al.}(2015)\citenamefont {Koochi}, \citenamefont {Farrokhabadi},\ and\ \citenamefont {Abadyan}}]{Koochi2015}%
  \BibitemOpen
  \bibfield  {author} {\bibinfo {author} {\bibfnamefont {A.}~\bibnamefont {Koochi}}, \bibinfo {author} {\bibfnamefont {A.}~\bibnamefont {Farrokhabadi}},\ and\ \bibinfo {author} {\bibfnamefont {M.}~\bibnamefont {Abadyan}},\ }\href {https://doi.org/10.1007/s00542-014-2183-y} {\bibfield  {journal} {\bibinfo  {journal} {Microsyst. Technol.}\ }\textbf {\bibinfo {volume} {21}},\ \bibinfo {pages} {355} (\bibinfo {year} {2015})}\BibitemShut {NoStop}%
\bibitem [{\citenamefont {Yazdanpanahi}\ \emph {et~al.}(2014)\citenamefont {Yazdanpanahi}, \citenamefont {Noghrehabadi},\ and\ \citenamefont {Ghalambaz}}]{Yazdanpanahi2014}%
  \BibitemOpen
  \bibfield  {author} {\bibinfo {author} {\bibfnamefont {E.}~\bibnamefont {Yazdanpanahi}}, \bibinfo {author} {\bibfnamefont {A.}~\bibnamefont {Noghrehabadi}},\ and\ \bibinfo {author} {\bibfnamefont {M.}~\bibnamefont {Ghalambaz}},\ }\href {https://doi.org/10.1016/j.ijnonlinmec.2013.09.001} {\bibfield  {journal} {\bibinfo  {journal} {Int. J. Non-Linear Mech.}\ }\textbf {\bibinfo {volume} {58}},\ \bibinfo {pages} {128} (\bibinfo {year} {2014})}\BibitemShut {NoStop}%
\bibitem [{\citenamefont {Farrokhabadi}\ \emph {et~al.}(2016)\citenamefont {Farrokhabadi}, \citenamefont {Mohebshahedin}, \citenamefont {Rach},\ and\ \citenamefont {Duan}}]{Farrokhabadi2016}%
  \BibitemOpen
  \bibfield  {author} {\bibinfo {author} {\bibfnamefont {A.}~\bibnamefont {Farrokhabadi}}, \bibinfo {author} {\bibfnamefont {A.}~\bibnamefont {Mohebshahedin}}, \bibinfo {author} {\bibfnamefont {R.}~\bibnamefont {Rach}},\ and\ \bibinfo {author} {\bibfnamefont {J.~S.}\ \bibnamefont {Duan}},\ }\href {https://doi.org/10.1016/j.physe.2015.09.033} {\bibfield  {journal} {\bibinfo  {journal} {Phys. E: Low-Dimens. Syst. Nanostructures}\ }\textbf {\bibinfo {volume} {75}},\ \bibinfo {pages} {202} (\bibinfo {year} {2016})}\BibitemShut {NoStop}%
\bibitem [{\citenamefont {Ouakad}(2017)}]{Ouakad2017}%
  \BibitemOpen
  \bibfield  {author} {\bibinfo {author} {\bibfnamefont {H.~M.}\ \bibnamefont {Ouakad}},\ }\href {https://doi.org/10.1007/s00542-017-3356-2} {\bibfield  {journal} {\bibinfo  {journal} {Microsystem Technologies}\ }\textbf {\bibinfo {volume} {23}},\ \bibinfo {pages} {5903} (\bibinfo {year} {2017})}\BibitemShut {NoStop}%
\bibitem [{\citenamefont {Liem}\ \emph {et~al.}(2020)\citenamefont {Liem}, \citenamefont {Ari}, \citenamefont {McDaniel},\ and\ \citenamefont {Ekinci}}]{liem2020inverse}%
  \BibitemOpen
  \bibfield  {author} {\bibinfo {author} {\bibfnamefont {A.~T.}\ \bibnamefont {Liem}}, \bibinfo {author} {\bibfnamefont {A.~B.}\ \bibnamefont {Ari}}, \bibinfo {author} {\bibfnamefont {J.~G.}\ \bibnamefont {McDaniel}},\ and\ \bibinfo {author} {\bibfnamefont {K.~L.}\ \bibnamefont {Ekinci}},\ }\href@noop {} {\bibfield  {journal} {\bibinfo  {journal} {J. Appl. Mech.}\ }\textbf {\bibinfo {volume} {87}},\ \bibinfo {pages} {061002} (\bibinfo {year} {2020})}\BibitemShut {NoStop}%
\bibitem [{\citenamefont {Barbish}\ \emph {et~al.}(2022)\citenamefont {Barbish}, \citenamefont {Ti}, \citenamefont {Ekinci},\ and\ \citenamefont {Paul}}]{Barbish2022}%
  \BibitemOpen
  \bibfield  {author} {\bibinfo {author} {\bibfnamefont {J.}~\bibnamefont {Barbish}}, \bibinfo {author} {\bibfnamefont {C.}~\bibnamefont {Ti}}, \bibinfo {author} {\bibfnamefont {K.}~\bibnamefont {Ekinci}},\ and\ \bibinfo {author} {\bibfnamefont {M.}~\bibnamefont {Paul}},\ }\href@noop {} {\bibfield  {journal} {\bibinfo  {journal} {J. Appl. Phys.}\ }\textbf {\bibinfo {volume} {132}},\ \bibinfo {pages} {034501} (\bibinfo {year} {2022})}\BibitemShut {NoStop}%
\bibitem [{\citenamefont {Hajjaj}\ \emph {et~al.}(2020)\citenamefont {Hajjaj}, \citenamefont {Jaber}, \citenamefont {Ilyas}, \citenamefont {Alfosail},\ and\ \citenamefont {Younis}}]{Hajjaj2020}%
  \BibitemOpen
  \bibfield  {author} {\bibinfo {author} {\bibfnamefont {A.}~\bibnamefont {Hajjaj}}, \bibinfo {author} {\bibfnamefont {N.}~\bibnamefont {Jaber}}, \bibinfo {author} {\bibfnamefont {S.}~\bibnamefont {Ilyas}}, \bibinfo {author} {\bibfnamefont {F.}~\bibnamefont {Alfosail}},\ and\ \bibinfo {author} {\bibfnamefont {M.~I.}\ \bibnamefont {Younis}},\ }\href@noop {} {\bibfield  {journal} {\bibinfo  {journal} {Int. J. Non-Linear Mech.}\ }\textbf {\bibinfo {volume} {119}},\ \bibinfo {pages} {103328} (\bibinfo {year} {2020})}\BibitemShut {NoStop}%
\bibitem [{\citenamefont {Khaniki}\ \emph {et~al.}(2021)\citenamefont {Khaniki}, \citenamefont {Ghayesh},\ and\ \citenamefont {Amabili}}]{Khaniki2021}%
  \BibitemOpen
  \bibfield  {author} {\bibinfo {author} {\bibfnamefont {H.~B.}\ \bibnamefont {Khaniki}}, \bibinfo {author} {\bibfnamefont {M.~H.}\ \bibnamefont {Ghayesh}},\ and\ \bibinfo {author} {\bibfnamefont {M.}~\bibnamefont {Amabili}},\ }\href@noop {} {\bibfield  {journal} {\bibinfo  {journal} {Int. J. Non-Linear Mech.}\ }\textbf {\bibinfo {volume} {129}},\ \bibinfo {pages} {103658} (\bibinfo {year} {2021})}\BibitemShut {NoStop}%
\bibitem [{\citenamefont {Masmanidis}\ \emph {et~al.}(2007)\citenamefont {Masmanidis}, \citenamefont {Karabalin}, \citenamefont {De~Vlaminck}, \citenamefont {Borghs}, \citenamefont {Freeman},\ and\ \citenamefont {Roukes}}]{Masmanidis2007}%
  \BibitemOpen
  \bibfield  {author} {\bibinfo {author} {\bibfnamefont {S.~C.}\ \bibnamefont {Masmanidis}}, \bibinfo {author} {\bibfnamefont {R.~B.}\ \bibnamefont {Karabalin}}, \bibinfo {author} {\bibfnamefont {I.}~\bibnamefont {De~Vlaminck}}, \bibinfo {author} {\bibfnamefont {G.}~\bibnamefont {Borghs}}, \bibinfo {author} {\bibfnamefont {M.~R.}\ \bibnamefont {Freeman}},\ and\ \bibinfo {author} {\bibfnamefont {M.~L.}\ \bibnamefont {Roukes}},\ }\href@noop {} {\bibfield  {journal} {\bibinfo  {journal} {Science}\ }\textbf {\bibinfo {volume} {317}},\ \bibinfo {pages} {780} (\bibinfo {year} {2007})}\BibitemShut {NoStop}%
\bibitem [{\citenamefont {Wasisto}\ \emph {et~al.}(2014)\citenamefont {Wasisto}, \citenamefont {Huang}, \citenamefont {Merzsch}, \citenamefont {Stranz}, \citenamefont {Waag},\ and\ \citenamefont {Peiner}}]{Wasisto2014}%
  \BibitemOpen
  \bibfield  {author} {\bibinfo {author} {\bibfnamefont {H.~S.}\ \bibnamefont {Wasisto}}, \bibinfo {author} {\bibfnamefont {K.}~\bibnamefont {Huang}}, \bibinfo {author} {\bibfnamefont {S.}~\bibnamefont {Merzsch}}, \bibinfo {author} {\bibfnamefont {A.}~\bibnamefont {Stranz}}, \bibinfo {author} {\bibfnamefont {A.}~\bibnamefont {Waag}},\ and\ \bibinfo {author} {\bibfnamefont {E.}~\bibnamefont {Peiner}},\ }\href@noop {} {\bibfield  {journal} {\bibinfo  {journal} {Microsyst. Technol.}\ }\textbf {\bibinfo {volume} {20}},\ \bibinfo {pages} {571} (\bibinfo {year} {2014})}\BibitemShut {NoStop}%
\bibitem [{\citenamefont {Truitt}\ \emph {et~al.}(2006)\citenamefont {Truitt}, \citenamefont {Hertzberg}, \citenamefont {Huang}, \citenamefont {Ekinci},\ and\ \citenamefont {Schwab}}]{truitt2006}%
  \BibitemOpen
  \bibfield  {author} {\bibinfo {author} {\bibfnamefont {P.~A.}\ \bibnamefont {Truitt}}, \bibinfo {author} {\bibfnamefont {J.~B.}\ \bibnamefont {Hertzberg}}, \bibinfo {author} {\bibfnamefont {C.~C.}\ \bibnamefont {Huang}}, \bibinfo {author} {\bibfnamefont {K.~L.}\ \bibnamefont {Ekinci}},\ and\ \bibinfo {author} {\bibfnamefont {K.~C.}\ \bibnamefont {Schwab}},\ }\href {https://doi.org/10.1021/nl062278g} {\bibfield  {journal} {\bibinfo  {journal} {Nano Lett.}\ }\textbf {\bibinfo {volume} {7}},\ \bibinfo {pages} {120} (\bibinfo {year} {2006})}\BibitemShut {NoStop}%
\bibitem [{\citenamefont {Jiang}\ \emph {et~al.}(2006)\citenamefont {Jiang}, \citenamefont {Cheung}, \citenamefont {Hedley}, \citenamefont {Hassan}, \citenamefont {Harris}, \citenamefont {Burdess}, \citenamefont {Mehregany},\ and\ \citenamefont {Zorman}}]{jiang2006sic}%
  \BibitemOpen
  \bibfield  {author} {\bibinfo {author} {\bibfnamefont {L.}~\bibnamefont {Jiang}}, \bibinfo {author} {\bibfnamefont {R.}~\bibnamefont {Cheung}}, \bibinfo {author} {\bibfnamefont {J.}~\bibnamefont {Hedley}}, \bibinfo {author} {\bibfnamefont {M.}~\bibnamefont {Hassan}}, \bibinfo {author} {\bibfnamefont {A.}~\bibnamefont {Harris}}, \bibinfo {author} {\bibfnamefont {J.}~\bibnamefont {Burdess}}, \bibinfo {author} {\bibfnamefont {M.}~\bibnamefont {Mehregany}},\ and\ \bibinfo {author} {\bibfnamefont {C.}~\bibnamefont {Zorman}},\ }\href@noop {} {\bibfield  {journal} {\bibinfo  {journal} {Sens. Actuator A Phys.}\ }\textbf {\bibinfo {volume} {128}},\ \bibinfo {pages} {376} (\bibinfo {year} {2006})}\BibitemShut {NoStop}%
\bibitem [{\citenamefont {Reichenbach}\ \emph {et~al.}(2006)\citenamefont {Reichenbach}, \citenamefont {Zalalutdinov}, \citenamefont {Parpia},\ and\ \citenamefont {Craighead}}]{reichenbach2006}%
  \BibitemOpen
  \bibfield  {author} {\bibinfo {author} {\bibfnamefont {R.}~\bibnamefont {Reichenbach}}, \bibinfo {author} {\bibfnamefont {M.}~\bibnamefont {Zalalutdinov}}, \bibinfo {author} {\bibfnamefont {J.}~\bibnamefont {Parpia}},\ and\ \bibinfo {author} {\bibfnamefont {H.}~\bibnamefont {Craighead}},\ }\href {https://doi.org/10.1109/led.2006.882526} {\bibfield  {journal} {\bibinfo  {journal} {IEEE Electron Device Lett.}\ }\textbf {\bibinfo {volume} {27}},\ \bibinfo {pages} {805} (\bibinfo {year} {2006})}\BibitemShut {NoStop}%
\bibitem [{\citenamefont {Ren}\ \emph {et~al.}(2021)\citenamefont {Ren}, \citenamefont {Yuan}, \citenamefont {Su}, \citenamefont {Mangla}, \citenamefont {Nam}, \citenamefont {Lu}, \citenamefont {Tenney},\ and\ \citenamefont {Shi}}]{ren2021electro}%
  \BibitemOpen
  \bibfield  {author} {\bibinfo {author} {\bibfnamefont {Z.}~\bibnamefont {Ren}}, \bibinfo {author} {\bibfnamefont {J.}~\bibnamefont {Yuan}}, \bibinfo {author} {\bibfnamefont {X.}~\bibnamefont {Su}}, \bibinfo {author} {\bibfnamefont {S.}~\bibnamefont {Mangla}}, \bibinfo {author} {\bibfnamefont {C.-Y.}\ \bibnamefont {Nam}}, \bibinfo {author} {\bibfnamefont {M.}~\bibnamefont {Lu}}, \bibinfo {author} {\bibfnamefont {S.~A.}\ \bibnamefont {Tenney}},\ and\ \bibinfo {author} {\bibfnamefont {Y.}~\bibnamefont {Shi}},\ }\href@noop {} {\bibfield  {journal} {\bibinfo  {journal} {Microsyst. Technol.}\ }\textbf {\bibinfo {volume} {27}},\ \bibinfo {pages} {2041} (\bibinfo {year} {2021})}\BibitemShut {NoStop}%
\bibitem [{\citenamefont {Cauchi}\ \emph {et~al.}(2018)\citenamefont {Cauchi}, \citenamefont {Grech}, \citenamefont {Mallia}, \citenamefont {Mollicone},\ and\ \citenamefont {Sammut}}]{cauchi2018analytical}%
  \BibitemOpen
  \bibfield  {author} {\bibinfo {author} {\bibfnamefont {M.}~\bibnamefont {Cauchi}}, \bibinfo {author} {\bibfnamefont {I.}~\bibnamefont {Grech}}, \bibinfo {author} {\bibfnamefont {B.}~\bibnamefont {Mallia}}, \bibinfo {author} {\bibfnamefont {P.}~\bibnamefont {Mollicone}},\ and\ \bibinfo {author} {\bibfnamefont {N.}~\bibnamefont {Sammut}},\ }\href@noop {} {\bibfield  {journal} {\bibinfo  {journal} {Micromachines}\ }\textbf {\bibinfo {volume} {9}},\ \bibinfo {pages} {108} (\bibinfo {year} {2018})}\BibitemShut {NoStop}%
\bibitem [{\citenamefont {Masood}\ \emph {et~al.}(2019)\citenamefont {Masood}, \citenamefont {Saleem}, \citenamefont {Khan},\ and\ \citenamefont {Hamza}}]{masood2019design}%
  \BibitemOpen
  \bibfield  {author} {\bibinfo {author} {\bibfnamefont {M.~U.}\ \bibnamefont {Masood}}, \bibinfo {author} {\bibfnamefont {M.~M.}\ \bibnamefont {Saleem}}, \bibinfo {author} {\bibfnamefont {U.~S.}\ \bibnamefont {Khan}},\ and\ \bibinfo {author} {\bibfnamefont {A.}~\bibnamefont {Hamza}},\ }\href@noop {} {\bibfield  {journal} {\bibinfo  {journal} {Microsyst. Technol.}\ }\textbf {\bibinfo {volume} {25}},\ \bibinfo {pages} {1171} (\bibinfo {year} {2019})}\BibitemShut {NoStop}%
\bibitem [{\citenamefont {Thangavel}\ \emph {et~al.}(2018)\citenamefont {Thangavel}, \citenamefont {Rengaswamy}, \citenamefont {Sukumar},\ and\ \citenamefont {Sekar}}]{thangavel2018modelling}%
  \BibitemOpen
  \bibfield  {author} {\bibinfo {author} {\bibfnamefont {A.}~\bibnamefont {Thangavel}}, \bibinfo {author} {\bibfnamefont {R.}~\bibnamefont {Rengaswamy}}, \bibinfo {author} {\bibfnamefont {P.~K.}\ \bibnamefont {Sukumar}},\ and\ \bibinfo {author} {\bibfnamefont {R.}~\bibnamefont {Sekar}},\ }\href@noop {} {\bibfield  {journal} {\bibinfo  {journal} {Microsyst. Technol.}\ }\textbf {\bibinfo {volume} {24}},\ \bibinfo {pages} {1767} (\bibinfo {year} {2018})}\BibitemShut {NoStop}%
\bibitem [{\citenamefont {Hussein}\ \emph {et~al.}(2020)\citenamefont {Hussein}, \citenamefont {Fariborzi},\ and\ \citenamefont {Younis}}]{hussein2020modeling}%
  \BibitemOpen
  \bibfield  {author} {\bibinfo {author} {\bibfnamefont {H.}~\bibnamefont {Hussein}}, \bibinfo {author} {\bibfnamefont {H.}~\bibnamefont {Fariborzi}},\ and\ \bibinfo {author} {\bibfnamefont {M.~I.}\ \bibnamefont {Younis}},\ }\href@noop {} {\bibfield  {journal} {\bibinfo  {journal} {J. Microelectromechanical Syst.}\ }\textbf {\bibinfo {volume} {29}},\ \bibinfo {pages} {1570} (\bibinfo {year} {2020})}\BibitemShut {NoStop}%
\bibitem [{\citenamefont {Kim}\ \emph {et~al.}(2022)\citenamefont {Kim}, \citenamefont {Kim},\ and\ \citenamefont {Kim}}]{kim2022analytical}%
  \BibitemOpen
  \bibfield  {author} {\bibinfo {author} {\bibfnamefont {S.}~\bibnamefont {Kim}}, \bibinfo {author} {\bibfnamefont {K.-S.}\ \bibnamefont {Kim}},\ and\ \bibinfo {author} {\bibfnamefont {Y.}~\bibnamefont {Kim}},\ }\href@noop {} {\bibfield  {journal} {\bibinfo  {journal} {Sens. Actuator A Phys.}\ }\textbf {\bibinfo {volume} {348}},\ \bibinfo {pages} {113984} (\bibinfo {year} {2022})}\BibitemShut {NoStop}%
\bibitem [{\citenamefont {Cleland}\ \emph {et~al.}(2001)\citenamefont {Cleland}, \citenamefont {Pophristic},\ and\ \citenamefont {Ferguson}}]{Cleland2001}%
  \BibitemOpen
  \bibfield  {author} {\bibinfo {author} {\bibfnamefont {A.}~\bibnamefont {Cleland}}, \bibinfo {author} {\bibfnamefont {M.}~\bibnamefont {Pophristic}},\ and\ \bibinfo {author} {\bibfnamefont {I.}~\bibnamefont {Ferguson}},\ }\href@noop {} {\bibfield  {journal} {\bibinfo  {journal} {Appl. Phys. Lett.}\ }\textbf {\bibinfo {volume} {79}},\ \bibinfo {pages} {2070} (\bibinfo {year} {2001})}\BibitemShut {NoStop}%
\bibitem [{\citenamefont {Knobel}\ and\ \citenamefont {Cleland}(2002)}]{Knobel2002}%
  \BibitemOpen
  \bibfield  {author} {\bibinfo {author} {\bibfnamefont {R.}~\bibnamefont {Knobel}}\ and\ \bibinfo {author} {\bibfnamefont {A.~N.}\ \bibnamefont {Cleland}},\ }\href {https://doi.org/10.1063/1.1507616} {\bibfield  {journal} {\bibinfo  {journal} {Appl. Phys. Lett.}\ }\textbf {\bibinfo {volume} {81}},\ \bibinfo {pages} {2258} (\bibinfo {year} {2002})}\BibitemShut {NoStop}%
\bibitem [{\citenamefont {Adachi}(1985)}]{Adachi1985}%
  \BibitemOpen
  \bibfield  {author} {\bibinfo {author} {\bibfnamefont {S.}~\bibnamefont {Adachi}},\ }\href@noop {} {\bibfield  {journal} {\bibinfo  {journal} {J. Appl. Phys.}\ }\textbf {\bibinfo {volume} {58}},\ \bibinfo {pages} {R1} (\bibinfo {year} {1985})}\BibitemShut {NoStop}%
\bibitem [{\citenamefont {Auld}(1973)}]{piezoelastic}%
  \BibitemOpen
  \bibfield  {author} {\bibinfo {author} {\bibfnamefont {B.~A.}\ \bibnamefont {Auld}},\ }\href@noop {} {\emph {\bibinfo {title} {Acoustic fields and waves in solids}}}\ (\bibinfo  {publisher} {Wiley},\ \bibinfo {year} {1973})\BibitemShut {NoStop}%
\bibitem [{\citenamefont {Sinha}\ \emph {et~al.}(2009)\citenamefont {Sinha}, \citenamefont {Wabiszewski}, \citenamefont {Mahameed}, \citenamefont {Felmetsger}, \citenamefont {Tanner}, \citenamefont {Carpick},\ and\ \citenamefont {Piazza}}]{Sinha2009}%
  \BibitemOpen
  \bibfield  {author} {\bibinfo {author} {\bibfnamefont {N.}~\bibnamefont {Sinha}}, \bibinfo {author} {\bibfnamefont {G.~E.}\ \bibnamefont {Wabiszewski}}, \bibinfo {author} {\bibfnamefont {R.}~\bibnamefont {Mahameed}}, \bibinfo {author} {\bibfnamefont {V.~V.}\ \bibnamefont {Felmetsger}}, \bibinfo {author} {\bibfnamefont {S.~M.}\ \bibnamefont {Tanner}}, \bibinfo {author} {\bibfnamefont {R.~W.}\ \bibnamefont {Carpick}},\ and\ \bibinfo {author} {\bibfnamefont {G.}~\bibnamefont {Piazza}},\ }\href@noop {} {\bibfield  {journal} {\bibinfo  {journal} {Appl. Phys. Lett.}\ }\textbf {\bibinfo {volume} {95}},\ \bibinfo {pages} {053106} (\bibinfo {year} {2009})}\BibitemShut {NoStop}%
\bibitem [{\citenamefont {Truitt}\ \emph {et~al.}(2007)\citenamefont {Truitt}, \citenamefont {Hertzberg}, \citenamefont {Huang}, \citenamefont {Ekinci},\ and\ \citenamefont {Schwab}}]{schwab2007capacitive}%
  \BibitemOpen
  \bibfield  {author} {\bibinfo {author} {\bibfnamefont {P.~A.}\ \bibnamefont {Truitt}}, \bibinfo {author} {\bibfnamefont {J.~B.}\ \bibnamefont {Hertzberg}}, \bibinfo {author} {\bibfnamefont {C.}~\bibnamefont {Huang}}, \bibinfo {author} {\bibfnamefont {K.~L.}\ \bibnamefont {Ekinci}},\ and\ \bibinfo {author} {\bibfnamefont {K.~C.}\ \bibnamefont {Schwab}},\ }\href@noop {} {\bibfield  {journal} {\bibinfo  {journal} {Nano Lett.}\ }\textbf {\bibinfo {volume} {7}},\ \bibinfo {pages} {120} (\bibinfo {year} {2007})}\BibitemShut {NoStop}%
\bibitem [{\citenamefont {Bao}(2005)}]{Bao2005MEMSbook}%
  \BibitemOpen
  \bibfield  {author} {\bibinfo {author} {\bibfnamefont {M.}~\bibnamefont {Bao}},\ }\href@noop {} {\emph {\bibinfo {title} {Analysis and design principles of MEMS devices}}}\ (\bibinfo  {publisher} {Elsevier},\ \bibinfo {year} {2005})\BibitemShut {NoStop}%
\bibitem [{\citenamefont {Younis}(2011)}]{younis2011}%
  \BibitemOpen
  \bibfield  {author} {\bibinfo {author} {\bibfnamefont {M.~I.}\ \bibnamefont {Younis}},\ }\href@noop {} {\emph {\bibinfo {title} {MEMS linear and nonlinear statics and dynamics}}}\ (\bibinfo  {publisher} {Springer},\ \bibinfo {year} {2011})\BibitemShut {NoStop}%
\bibitem [{\citenamefont {Shavezipur}\ \emph {et~al.}(2011)\citenamefont {Shavezipur}, \citenamefont {Li}, \citenamefont {Laboriante}, \citenamefont {Gou}, \citenamefont {Carraro},\ and\ \citenamefont {Maboudian}}]{Shavezipur2011}%
  \BibitemOpen
  \bibfield  {author} {\bibinfo {author} {\bibfnamefont {M.}~\bibnamefont {Shavezipur}}, \bibinfo {author} {\bibfnamefont {G.~H.}\ \bibnamefont {Li}}, \bibinfo {author} {\bibfnamefont {I.}~\bibnamefont {Laboriante}}, \bibinfo {author} {\bibfnamefont {W.~J.}\ \bibnamefont {Gou}}, \bibinfo {author} {\bibfnamefont {C.}~\bibnamefont {Carraro}},\ and\ \bibinfo {author} {\bibfnamefont {R.}~\bibnamefont {Maboudian}},\ }\href {https://doi.org/10.1088/0960-1317/21/11/115025} {\bibfield  {journal} {\bibinfo  {journal} {J. Micromech. Microeng.}\ }\textbf {\bibinfo {volume} {21}},\ \bibinfo {pages} {115025} (\bibinfo {year} {2011})}\BibitemShut {NoStop}%
\bibitem [{\citenamefont {Mao}\ \emph {et~al.}(2010)\citenamefont {Mao}, \citenamefont {Wang}, \citenamefont {Wu}, \citenamefont {Tang},\ and\ \citenamefont {Ding}}]{mao2010}%
  \BibitemOpen
  \bibfield  {author} {\bibinfo {author} {\bibfnamefont {S.}~\bibnamefont {Mao}}, \bibinfo {author} {\bibfnamefont {H.}~\bibnamefont {Wang}}, \bibinfo {author} {\bibfnamefont {Y.}~\bibnamefont {Wu}}, \bibinfo {author} {\bibfnamefont {J.}~\bibnamefont {Tang}},\ and\ \bibinfo {author} {\bibfnamefont {G.}~\bibnamefont {Ding}},\ }in\ \href@noop {} {\emph {\bibinfo {booktitle} {2010 IEEE 5th International Conference on Nano/Micro Engineered and Molecular Systems}}}\ (\bibinfo {organization} {IEEE},\ \bibinfo {year} {2010})\ pp.\ \bibinfo {pages} {732--735}\BibitemShut {NoStop}%
\bibitem [{\citenamefont {Zhu}\ and\ \citenamefont {Pal}(2021)}]{zhu2021}%
  \BibitemOpen
  \bibfield  {author} {\bibinfo {author} {\bibfnamefont {Y.}~\bibnamefont {Zhu}}\ and\ \bibinfo {author} {\bibfnamefont {J.}~\bibnamefont {Pal}},\ }\href {https://doi.org/10.3390/mi12101237} {\bibfield  {journal} {\bibinfo  {journal} {Micromachines}\ }\textbf {\bibinfo {volume} {12}},\ \bibinfo {pages} {1237} (\bibinfo {year} {2021})}\BibitemShut {NoStop}%
\bibitem [{\citenamefont {Goldsmith}\ \emph {et~al.}(1999)\citenamefont {Goldsmith}, \citenamefont {Malczewski}, \citenamefont {Yao}, \citenamefont {Chen}, \citenamefont {Ehmke},\ and\ \citenamefont {Hinzel}}]{goldsmith1999}%
  \BibitemOpen
  \bibfield  {author} {\bibinfo {author} {\bibfnamefont {C.~L.}\ \bibnamefont {Goldsmith}}, \bibinfo {author} {\bibfnamefont {A.}~\bibnamefont {Malczewski}}, \bibinfo {author} {\bibfnamefont {Z.~J.}\ \bibnamefont {Yao}}, \bibinfo {author} {\bibfnamefont {S.}~\bibnamefont {Chen}}, \bibinfo {author} {\bibfnamefont {J.}~\bibnamefont {Ehmke}},\ and\ \bibinfo {author} {\bibfnamefont {D.~H.}\ \bibnamefont {Hinzel}},\ }\href {https://doi.org/10.1002/(sici)1099-047x(199907)9:4<362::aid-mmce7>3.0.co;2-h} {\bibfield  {journal} {\bibinfo  {journal} {Int. J. RF Microw. Comput.-Aided Eng.}\ }\textbf {\bibinfo {volume} {9}},\ \bibinfo {pages} {362} (\bibinfo {year} {1999})}\BibitemShut {NoStop}%
\bibitem [{\citenamefont {Parameswaran}\ \emph {et~al.}(1990)\citenamefont {Parameswaran}, \citenamefont {Ristic}, \citenamefont {Chau}, \citenamefont {Robinson},\ and\ \citenamefont {Allegretto}}]{parameswaran1990cmos}%
  \BibitemOpen
  \bibfield  {author} {\bibinfo {author} {\bibfnamefont {M.}~\bibnamefont {Parameswaran}}, \bibinfo {author} {\bibfnamefont {L.}~\bibnamefont {Ristic}}, \bibinfo {author} {\bibfnamefont {K.}~\bibnamefont {Chau}}, \bibinfo {author} {\bibfnamefont {A.}~\bibnamefont {Robinson}},\ and\ \bibinfo {author} {\bibfnamefont {W.}~\bibnamefont {Allegretto}},\ }in\ \href@noop {} {\emph {\bibinfo {booktitle} {IEEE Proceedings on Micro Electro Mechanical Systems, An Investigation of Micro Structures, Sensors, Actuators, Machines and Robots.}}}\ (\bibinfo {organization} {IEEE},\ \bibinfo {year} {1990})\ pp.\ \bibinfo {pages} {128--131}\BibitemShut {NoStop}%
\bibitem [{\citenamefont {Reinke}\ \emph {et~al.}(2010)\citenamefont {Reinke}, \citenamefont {Fedder},\ and\ \citenamefont {Mukherjee}}]{reinke2010}%
  \BibitemOpen
  \bibfield  {author} {\bibinfo {author} {\bibfnamefont {J.}~\bibnamefont {Reinke}}, \bibinfo {author} {\bibfnamefont {G.~K.}\ \bibnamefont {Fedder}},\ and\ \bibinfo {author} {\bibfnamefont {T.}~\bibnamefont {Mukherjee}},\ }\href@noop {} {\bibfield  {journal} {\bibinfo  {journal} {J. Microelectromechanical Syst.}\ }\textbf {\bibinfo {volume} {19}},\ \bibinfo {pages} {1105} (\bibinfo {year} {2010})}\BibitemShut {NoStop}%
\bibitem [{\citenamefont {Lara-Castro}\ \emph {et~al.}(2017)\citenamefont {Lara-Castro}, \citenamefont {Herrera-Amaya}, \citenamefont {Escarola-Rosas}, \citenamefont {V{\'a}zquez-Toledo}, \citenamefont {L{\'o}pez-Huerta}, \citenamefont {Aguilera-Cort{\'e}s},\ and\ \citenamefont {Herrera-May}}]{Castro2017}%
  \BibitemOpen
  \bibfield  {author} {\bibinfo {author} {\bibfnamefont {M.}~\bibnamefont {Lara-Castro}}, \bibinfo {author} {\bibfnamefont {A.}~\bibnamefont {Herrera-Amaya}}, \bibinfo {author} {\bibfnamefont {M.~A.}\ \bibnamefont {Escarola-Rosas}}, \bibinfo {author} {\bibfnamefont {M.}~\bibnamefont {V{\'a}zquez-Toledo}}, \bibinfo {author} {\bibfnamefont {F.}~\bibnamefont {L{\'o}pez-Huerta}}, \bibinfo {author} {\bibfnamefont {L.~A.}\ \bibnamefont {Aguilera-Cort{\'e}s}},\ and\ \bibinfo {author} {\bibfnamefont {A.~L.}\ \bibnamefont {Herrera-May}},\ }\href@noop {} {\bibfield  {journal} {\bibinfo  {journal} {Micromachines}\ }\textbf {\bibinfo {volume} {8}},\ \bibinfo {pages} {203} (\bibinfo {year} {2017})}\BibitemShut {NoStop}%
\bibitem [{\citenamefont {Elbuken}\ \emph {et~al.}(2009)\citenamefont {Elbuken}, \citenamefont {Topaloglu}, \citenamefont {Nieva}, \citenamefont {Yavuz},\ and\ \citenamefont {Huissoon}}]{Elbuken2009}%
  \BibitemOpen
  \bibfield  {author} {\bibinfo {author} {\bibfnamefont {C.}~\bibnamefont {Elbuken}}, \bibinfo {author} {\bibfnamefont {N.}~\bibnamefont {Topaloglu}}, \bibinfo {author} {\bibfnamefont {P.~M.}\ \bibnamefont {Nieva}}, \bibinfo {author} {\bibfnamefont {M.}~\bibnamefont {Yavuz}},\ and\ \bibinfo {author} {\bibfnamefont {J.~P.}\ \bibnamefont {Huissoon}},\ }\href@noop {} {\bibfield  {journal} {\bibinfo  {journal} {Microsyst. Technol.}\ }\textbf {\bibinfo {volume} {15}},\ \bibinfo {pages} {713} (\bibinfo {year} {2009})}\BibitemShut {NoStop}%
\bibitem [{\citenamefont {Yan}\ \emph {et~al.}(2003)\citenamefont {Yan}, \citenamefont {Khajepour},\ and\ \citenamefont {Mansour}}]{Yan2003}%
  \BibitemOpen
  \bibfield  {author} {\bibinfo {author} {\bibfnamefont {D.}~\bibnamefont {Yan}}, \bibinfo {author} {\bibfnamefont {A.}~\bibnamefont {Khajepour}},\ and\ \bibinfo {author} {\bibfnamefont {R.}~\bibnamefont {Mansour}},\ }\href@noop {} {\bibfield  {journal} {\bibinfo  {journal} {J. Micromech. Microeng.}\ }\textbf {\bibinfo {volume} {13}},\ \bibinfo {pages} {312} (\bibinfo {year} {2003})}\BibitemShut {NoStop}%
\bibitem [{\citenamefont {Huang}\ and\ \citenamefont {Lee}(1999)}]{Huang1999}%
  \BibitemOpen
  \bibfield  {author} {\bibinfo {author} {\bibfnamefont {Q.-A.}\ \bibnamefont {Huang}}\ and\ \bibinfo {author} {\bibfnamefont {N.~K.~S.}\ \bibnamefont {Lee}},\ }\href@noop {} {\bibfield  {journal} {\bibinfo  {journal} {J. Micromech. Microeng.}\ }\textbf {\bibinfo {volume} {9}},\ \bibinfo {pages} {64} (\bibinfo {year} {1999})}\BibitemShut {NoStop}%
\bibitem [{\citenamefont {Shan}\ \emph {et~al.}(2017)\citenamefont {Shan}, \citenamefont {Qi}, \citenamefont {Cui},\ and\ \citenamefont {Zhou}}]{Shan2017}%
  \BibitemOpen
  \bibfield  {author} {\bibinfo {author} {\bibfnamefont {T.}~\bibnamefont {Shan}}, \bibinfo {author} {\bibfnamefont {X.}~\bibnamefont {Qi}}, \bibinfo {author} {\bibfnamefont {L.}~\bibnamefont {Cui}},\ and\ \bibinfo {author} {\bibfnamefont {X.}~\bibnamefont {Zhou}},\ }\href@noop {} {\bibfield  {journal} {\bibinfo  {journal} {Microsyst. Technol.}\ }\textbf {\bibinfo {volume} {23}},\ \bibinfo {pages} {2629} (\bibinfo {year} {2017})}\BibitemShut {NoStop}%
\bibitem [{\citenamefont {Enikov}\ \emph {et~al.}(2005)\citenamefont {Enikov}, \citenamefont {Kedar},\ and\ \citenamefont {Lazarov}}]{Enikov2005}%
  \BibitemOpen
  \bibfield  {author} {\bibinfo {author} {\bibfnamefont {E.~T.}\ \bibnamefont {Enikov}}, \bibinfo {author} {\bibfnamefont {S.~S.}\ \bibnamefont {Kedar}},\ and\ \bibinfo {author} {\bibfnamefont {K.~V.}\ \bibnamefont {Lazarov}},\ }\href@noop {} {\bibfield  {journal} {\bibinfo  {journal} {J. Microelectromechanical Syst.}\ }\textbf {\bibinfo {volume} {14}},\ \bibinfo {pages} {788} (\bibinfo {year} {2005})}\BibitemShut {NoStop}%
\bibitem [{\citenamefont {Hickey}\ \emph {et~al.}(2003)\citenamefont {Hickey}, \citenamefont {Sameoto}, \citenamefont {Hubbard},\ and\ \citenamefont {Kujath}}]{Hickey2003}%
  \BibitemOpen
  \bibfield  {author} {\bibinfo {author} {\bibfnamefont {R.}~\bibnamefont {Hickey}}, \bibinfo {author} {\bibfnamefont {D.}~\bibnamefont {Sameoto}}, \bibinfo {author} {\bibfnamefont {T.}~\bibnamefont {Hubbard}},\ and\ \bibinfo {author} {\bibfnamefont {M.}~\bibnamefont {Kujath}},\ }\href {https://doi.org/10.1088/0960-1317/13/1/306} {\bibfield  {journal} {\bibinfo  {journal} {J. Micromech. Microeng.}\ }\textbf {\bibinfo {volume} {13}},\ \bibinfo {pages} {40} (\bibinfo {year} {2003})}\BibitemShut {NoStop}%
\bibitem [{\citenamefont {Ti}\ \emph {et~al.}(2021)\citenamefont {Ti}, \citenamefont {Ari}, \citenamefont {Karakan}, \citenamefont {Yanik}, \citenamefont {Kaya}, \citenamefont {Hanay}, \citenamefont {Svitelskiy}, \citenamefont {Gonz{\'a}lez}, \citenamefont {Seren},\ and\ \citenamefont {Ekinci}}]{chaoyang2021}%
  \BibitemOpen
  \bibfield  {author} {\bibinfo {author} {\bibfnamefont {C.}~\bibnamefont {Ti}}, \bibinfo {author} {\bibfnamefont {A.~B.}\ \bibnamefont {Ari}}, \bibinfo {author} {\bibfnamefont {M.~C.}\ \bibnamefont {Karakan}}, \bibinfo {author} {\bibfnamefont {C.}~\bibnamefont {Yanik}}, \bibinfo {author} {\bibfnamefont {I.~I.}\ \bibnamefont {Kaya}}, \bibinfo {author} {\bibfnamefont {M.~S.}\ \bibnamefont {Hanay}}, \bibinfo {author} {\bibfnamefont {O.}~\bibnamefont {Svitelskiy}}, \bibinfo {author} {\bibfnamefont {M.}~\bibnamefont {Gonz{\'a}lez}}, \bibinfo {author} {\bibfnamefont {H.}~\bibnamefont {Seren}},\ and\ \bibinfo {author} {\bibfnamefont {K.~L.}\ \bibnamefont {Ekinci}},\ }\href@noop {} {\bibfield  {journal} {\bibinfo  {journal} {Nano Lett.}\ }\textbf {\bibinfo {volume} {21}},\ \bibinfo {pages} {6533} (\bibinfo {year} {2021})}\BibitemShut {NoStop}%
\bibitem [{\citenamefont {Ti}(2022)}]{Ticomm}%
  \BibitemOpen
  \bibfield  {author} {\bibinfo {author} {\bibfnamefont {C.}~\bibnamefont {Ti}},\ }\href@noop {} {}\bibinfo {howpublished} {personal communication} (\bibinfo {year} {2022})\BibitemShut {NoStop}%
\bibitem [{\citenamefont {Imboden}\ and\ \citenamefont {Mohanty}(2014)}]{imboden2014dissipation}%
  \BibitemOpen
  \bibfield  {author} {\bibinfo {author} {\bibfnamefont {M.}~\bibnamefont {Imboden}}\ and\ \bibinfo {author} {\bibfnamefont {P.}~\bibnamefont {Mohanty}},\ }\href@noop {} {\bibfield  {journal} {\bibinfo  {journal} {Phys. Rep.}\ }\textbf {\bibinfo {volume} {534}},\ \bibinfo {pages} {89} (\bibinfo {year} {2014})}\BibitemShut {NoStop}%
\bibitem [{\citenamefont {Kara}\ \emph {et~al.}(2017)\citenamefont {Kara}, \citenamefont {Yakhot},\ and\ \citenamefont {Ekinci}}]{kara2017generalized}%
  \BibitemOpen
  \bibfield  {author} {\bibinfo {author} {\bibfnamefont {V.}~\bibnamefont {Kara}}, \bibinfo {author} {\bibfnamefont {V.}~\bibnamefont {Yakhot}},\ and\ \bibinfo {author} {\bibfnamefont {K.~L.}\ \bibnamefont {Ekinci}},\ }\href@noop {} {\bibfield  {journal} {\bibinfo  {journal} {Phys. Rev. Lett.}\ }\textbf {\bibinfo {volume} {118}},\ \bibinfo {pages} {074505} (\bibinfo {year} {2017})}\BibitemShut {NoStop}%
\bibitem [{\citenamefont {{COMSOL Multiphysics v.6.0., COMSOL AB, Stockholm, Sweden}}(2022)}]{COMSOL}%
  \BibitemOpen
  \bibfield  {author} {\bibinfo {author} {\bibnamefont {{COMSOL Multiphysics v.6.0., COMSOL AB, Stockholm, Sweden}}},\ }\href {https://www.comsol.com/} {\bibinfo {title} {see www.comsol.com}} (\bibinfo {year} {2022})\BibitemShut {NoStop}%
\bibitem [{com(2021)}]{comsolsmmanual}%
  \BibitemOpen
  \href@noop {} {\emph {\bibinfo {title} {Structural Mechanics Module User’s Guide}}},\ \bibinfo {organization} {COMSOL Multiphysics v.6.0, COMSOL AB},\ \bibinfo {address} {Stockholm, Sweden} (\bibinfo {year} {2021})\BibitemShut {NoStop}%
\bibitem [{\citenamefont {Fomenkov}(2017)}]{fomenkov2017euv}%
  \BibitemOpen
  \bibfield  {author} {\bibinfo {author} {\bibfnamefont {I.}~\bibnamefont {Fomenkov}},\ }in\ \href@noop {} {\emph {\bibinfo {booktitle} {International Workshop on EUV Lithography}}}\ (\bibinfo {year} {2017})\BibitemShut {NoStop}%
\end{thebibliography}%
\end{document}